\documentclass{article}
\usepackage{graphicx} 
\usepackage{amssymb,amsmath}
\usepackage{booktabs}
\usepackage{siunitx}
\usepackage[nice]{nicefrac}

\newcommand{\dy}{\partial}
\newcommand{\dr}{\Delta_\rho}
\newcommand{\ds}{\Delta_s}
\newcommand{\dph}{\Delta_\phi}
\newcommand{\nr}{n_\rho}
\newcommand{\ns}{n_s}
\newcommand{\nph}{n_\phi}
\newcommand{\half}{\nicefrac{1}{2}}
\newcommand{\thr}{\nicefrac{3}{2}}
\newcommand{\Ab}{\boldsymbol{A}}
\newcommand{\Bb}{\boldsymbol{B}}

\newcommand{\jb}{\boldsymbol{j}}

\newcommand{\ex}{\,\mathrm{e}}
\newcommand{\evr}{\,\mathrm{e}_\rho}
\newcommand{\evs}{\,\mathrm{e}_s}
\newcommand{\evp}{\,\mathrm{e}_\phi}

\usepackage{geometry}
\title{Evaluation of Coronal and Interplanetary Magnetic Field Extrapolation Using PSP Solar Wind Observation}
\author{Yue-Chun Song}
\date{May 2023}
\begin{document}
\maketitle
\abstract{Using solar wind observation near PSP perihelions as constraints, we have investigated the parameters in various PFSS model methods. It's found that the interplanetary magnetic field extrapolation with source surface height $R_\mathrm{SS} = 2\,Rs$ is better than that with $R_\mathrm{SS} = 2.5\,Rs$. HMI and GONG magnetograms show similar performance in the simulation of magnetic field variation, but the former appears to have a slight advantage in reconstruction of intensity while the latter is more adaptable to sparser grids. The finite-difference method of constructing eigenvalue problem for potential field can achieve similar accuracy as analytic method and greatly improve the computational efficiency. MHD modeling performs relatively less well in magnetic field prediction, but it is able to provide rich information about solar-terrestrial space.
}
\section{Introduction}
Solar magnetic field is closely related to various structures and activities in solar-terrestrial space, and also an important factor affecting space weather. 
The dynamically changing magnetic field is the source of nearly all solar activity affecting earth and human technological systems. 
Yet due to the limited observational techniques, at present, only in-situ magnetic field measurement of spacecrafts and radial magnetic field measurement of photosphere are relatively accurate.
The observed chromospheric magnetic field has been continuously improved, while the direct measurement of the coronal magnetic field is still a difficult problem in solar physics \cite{YangZH2020b, YangZH2020a}. 

Some latest research has obtained the coronal magnetic field intensity distribution through indirect ways \cite{YangZH2020a}, but the commonly used method is still by extrapolation based on the measured photospheric magnetic field \cite{Wiegelmann2021}.
For certain coronal region where plasma $\beta\ll 1$, under the assumption of force-free field model, the Lorentz force is 0, that is,
\begin{equation}
     \jb \times \Bb = 0,
\end{equation}
\begin{equation}
     \nabla \times \Bb = \alpha \Bb.
     \label{equ:forcefree}
\end{equation}
Utilizing Maxwell's equations, that can be further simplified to a form containing only the magnetic field $\Bb$. If $\alpha$ is constant among that spatial range, $\Bb$ will be a linear force-free field, and particularly, a potential field where's no current with $\alpha=0$; if $\alpha$ varies in space, a nonlinear force-free field will be obtained.
When the premise $\beta\ll 1$ doesn't hold, a more comprehensive model is needed to calculate the coronal magnetic field, such as the magnetohydrostatic model \cite{Ruan2008}, the stationary magnetohydrodynamic model \cite{Wiegelmann2020}, magnetohydrodynamic model \cite{Mikic2018}, etc.

And for interplanetary space farther from the sun, the magnetic field can often be thought of as coupled to the plasma. As the solar wind spreads radially outward, a spiral structure is formed. That helical structure is also commonly referred to as "Parker spiral" due to Parker's seminal work on interplanetary magnetic field \cite{Parker1958}. Then at a larger distance from the heliocentric ($r \gg Rs$), $B_r \propto 1/r^2$, and $B_{\phi} \propto 1/r$.

Magnetometers are usually used to obtain the three-component magnetic field. We track the measurements at the satellite location to acquire the structure of local magnetic field. 
Before the launch of the Parker Solar Probe in 2018, the interplanetary magnetic field was mainly observed near 1 AU. 
PSP is able to reach the corona at about 9.5 solar radii from the sun and conduct direct measurements of the velocity of protons within 0.5 AU as well as the coronal and interplanetary magnetic field along its orbit, which provides more accurate solar wind speed input and new effective reference for optimizing the models.

In this work we mainly analyze the coronal and interplanetary magnetic field with Potential Field Source Surface (PFSS) model and magnetohydrodynamic (MHD) model.
Section 2 introduces the observational data used in this paper.
Section 3 describes several algorithms for PFSS model and MHD model in detail.
Section 4 is about the magnetic field simulation results and the comparison with in-situ observation. The adjustment of parameters and the selection of magnetograms are discussed.
Section 5 integrates the main conclusion and issues that still need further research.
\section{Data}
PSP (Parker Solar Probe) mission \cite{Fox2016} is to track how energy and heat are transported in solar corona, and to explore what drives the acceleration of solar wind and solar energetic particles. 
We focus on two groups of instruments here.
FIELDS (The electromagnetic fields investigation) captures the magnitude and direction of electric and magnetic fields in the solar atmosphere, and measures waves and turbulence in the inner heliosphere with high temporal resolution to understand the magnetic fields associated with waves, shock waves and magnetic reconnection, as well as electric fields in a wide frequency range measured directly or from a long distance \cite{Bale2016}.
SWEAP (The Solar Wind Electrons Alphas and Protons investigation) counts the richest particles (electrons, protons and helium ions) in the solar wind and measures properties such as velocity, density and temperature to improve our understanding of solar wind and coronal plasma \cite{Kasper2016}.
The radial component of proton velocity from SWEAP was used in this study as the observed data of solar wind velocity, while some simulation results were compared with the in-situ magnetic field measurements from FIELDS.

GONG (The Global Oscillation Network Group) aims to use helioseismology to conduct detailed research on the internal structure and dynamics of the sun. It relies on a network of six stations around the earth to achieve near-continuous observation. The synoptic map generated by GONG's zero point corrected magnetogram is used in the simulations in this paper, which corrects the zero-point uncertainty caused by heterogeneity and small imperfections in the magnetogram modulator on the basis of the standard magnetogram. Also, the polar field correction was carried out according to the lower-latitude observed magnetic field by a cubic polynomial surface fit \cite{Li2021}.

HMI (Helioseismic and Magnetic Imager) is one of the three instruments of SDO (Solar Dynamics Observatory), the main goal is to study the origin of solar changes and understand the internal structure of the sun as well as the various components of magnetic activity. HMI observes the motion of photosphere to study solar oscillation, studies the three components of the photospheric magnetic field according to the polarization of specific spectral lines, and makes high-resolution measurements of the vector magnetic field on the entire visible sun surface (SDO, HMI). Only synoptic maps for different Carrington Rotations (CR) are used here, which are made of magnetograms near the central meridian with a resolution of $3600\times 1440$.

K-COR (COSMO K-CORONAGRAPH) is one of the constituent instruments of the COSMO (The COronal Solar Magnetism Observatory) facility suite, dedicated to the study of the formation and dynamics of coronal mass ejections and the evolution of inner coronal density structure, which records the polarization brightness of light emitted by photosphere and scattered by free electrons in the corona. This paper uses K-COR observations as reference to verify the simulation of coronal magnetic field structure. The high-contrast K-COR white light image can clearly show the positions of various coronal structures.

We also evaluated some models with OMNI data set, which is primarily a 1963-to-current compilation of hourly-averaged, near-Earth solar wind magnetic field and plasma parameter data from several spacecraft in geocentric or L1 (Lagrange point) orbits.
\section{Physical models and calculation methods of coronal and interplanetary magnetic field}
\subsection{PFSS Model and Parker Spiral Field}
Potential Field Source Surface (PFSS) model \cite{Pfss} is widely used for magnetic field extrapolation from solar photosphere to corona and interplanetary space. Usually, the potential field solution within the designated source surface is obtained from the synoptic map by spherical harmonic function or finite difference method, and the magnetic field is extrapolated to interplanetary space considering the consistency between trajectory of solar wind and magnetic field structure. 

With the in-situ solar wind velocity measured by PSP, this helical magnetic connectivity can be expressed as
\[\phi(r)=\phi_0-\frac{\Omega}{V_{SW}}(r-r_0),\]
where $\phi(r),r$ are the longitude of a point at interplanetary space and its distance to solar center, $\Omega$ is solar rotation rate. And $\phi_0, r_0, V_\mathrm{SW}$ are the longitude, heliocentric radius, the radial solar wind speed observation of PSP respectively \cite{Badman}.
Then for each point at PSP trajectory, we can find the corresponding points at source surface or further interplanetary space along the magnetic flux tube it's located, with extrapolating the solar wind speed around those positions.

From source surface to interplanetary space, we use Parker spiral to describe the magnetic field \cite{Parker1958}. Source surface is assumed to have only $B_r$ components, while $B_\theta$ and $B_\phi$ components are zero. Since the magnetic field is coupled to plasma, the interplanetary magnetic field is actually governed by velocity field. Components of the velocity field can be expressed by following equations:
\begin{align}
    &v_r(r,\theta,\phi)=V_\mathrm{SW},\\
    &v_\theta(r,\theta,\phi)=0,\\
    &v_\phi(r,\theta,\phi)= \Omega (r-R_{ss})\sin\theta.
\end{align}
The corresponding components of magnetic field are:
\begin{align}
    &B_r(r,\theta,\phi)=B_r(R_{SS},\theta,\phi_0)(R_{SS}/r)^2,\\
    &B_\theta(r,\theta,\phi)=0,\\
    &B_\phi(r,\theta,\phi)=B_r(r,\theta,\phi)(\Omega / V_\mathrm{SW})(r-R_{SS})\sin\theta.
\end{align}

Magnetic field within the spherical shell region from photosphere to source surface can be solved by the following properties and boundary conditions:
\begin{align}
&\nabla\times\Bb=0\label{e00},\\
&\nabla\cdot\Bb=0\label{e01},\\
&B_r|_{r=1}=M(\theta,\phi),\\
&B_\theta|_{r=R_{SS}}=B_\phi|_{r=R_{SS}}=0,
\end{align}
where $M(\theta,\phi)$ represents the photospheric magnetic field measurement.
\subsubsection{Analytical solution of potential field in the form of spherical harmonics \cite{Linff}}
According to the irrotational property Equation \eqref{e00}, the scalar potential of $\Bb$ can be constructed so that
\[\Bb=\nabla\Phi,\]
At this time, Equation \eqref{e00} has been automatically satisfied, and it is only necessary to solve Equation \eqref{e01} under boundary conditions, namely
\[\Delta\Phi=0.\]
It has an analytical solution in spherical coordinate $(r,\theta,\phi)$:
\[\Phi(r,\theta,\phi)=\sum_{l=0}^{\infty}\sum_{m=-l}^l[A_{lm}r^l+b_{lm}r ^{-(l+1)}]Y_{lm}(\theta,\phi),\]
where $Y_{lm}$ is spherical harmonic function, and $A_{lm}$ and $B_{lm}$ are coefficients obtained according to boundary conditions.
Then the three components of magnetic field can be expressed as
\begin{align}
&B_r(r,\theta,\phi)=\sum_{l=0}^\infty\sum_{m=-l}^l[A_{lm}lr^{(l-1)}-B_{lm} (l+1)r^{-(l+2)}]Y_{lm}(\theta,\phi),\\
&B_\theta(r,\theta,\phi)=\frac1r\frac{\dy\Phi(r,\theta,\phi)}{\dy\theta},\\
&B_\phi(r,\theta,\phi)=\frac1{r\sin(\theta)}\frac{\dy\Phi(r,\theta,\phi)}{\dy\phi}.
\end{align}
Given the photospheric magnetic field $M(\theta,\phi)$, its spherical harmonic expansion can be written as
\begin{align}
&M(\theta,\phi)=\sum_{l=0}^{\infty}\sum_{m=-l}^{l}C_{lm}Y_{lm}(\theta,\phi),\ \
&C_{lm}=\int_0^{2\pi}\int_0^{\pi}Y_{lm}^*(\theta,\phi)M(\theta,\phi)\sin(\theta) d\theta d\phi,
\end{align}
where $Y_{lm}^*=(-1)^mY_{l,-m}$.
Based on von Neumann condition $B_r(r_0,\theta,\phi)=\frac{\dy\Phi}{\dy r}$ on photosphere we get
\[A_{lm}lr_0^{(l-1)}-B_{lm}(l+1)r_0^{-(l+2)}=C_{lm}.\]
Then according to the magnetic field turns radial at source surface $r=r_1$, that is, $B_\theta=B_\phi=0$, we have
\[A_{lm}r^l+b_{lm}r^{-(l+1)}=0.\]
So coefficients in the analytical solution of potential field will be
\[A_{lm}=\frac{C_{lm}r_0^{2+l}}{r_1^{1+2l}+l(r_0^{1+2l}+r_1^{1+2l})} ,\]
\[B_{lm}=-\frac{C_{lm}r_0^{2+l}r_1^{1+2l}}{r_1^{1+2l}+l(r_0^{1+2l}+r_1 ^{1+2l})}.\]
But infinite series cannot be calculated directly in practical application, only a limited number of $l$ can be intercepted for summation and the result is still an approximate solution.
\subsubsection{Finite Difference Iterative Potential-field Solver (\texttt{FDIPS}) \cite{Toth}}
Finite Difference Iterative Potential-field Solver (FDIPS) also transforms the problem into the solution of Laplace equation under boundary conditions by constructing a scalar potential. Under spherical coordinates $(r,\theta,\phi)$, $r,\cos\theta,\phi$ are evenly divided into $N_r,N_\theta,N_\phi$ units respectively. The magnetic field is discretized to cell faces and the scalar potential is discretized at cell centers, with an additional layer of ghost cells to represent the boundary conditions. Set cell centers as $(r_i,\theta_j,\phi_k)$, the magnetic field can be expressed in discrete gradient form as
\begin{align}
&B_{r,i+\half,j,k}=\frac{\Phi_{i+1,j,k}-\Phi_{i,j,k}}{\Delta r},\\
&B_{\theta,i,j+\half,k}=\frac{\sin\theta_{j+\half}(\Phi_{i,j+1,k}-\Phi_{i,j,k})} {r_i\Delta\cos\theta},\\
&B_{\phi,i,j,k+\half}=\frac{\Phi_{i,j,k+1}-\Phi_{i,j,k}}{r_i\sin\theta_j\Delta\phi} .
\end{align}\\
Then the divergence of magnetic field $\nabla^2\Phi$ can be approximated as
\begin{align}
0=(\nabla^2\Phi)_{i,j,k}=&\frac{r_{i+\half}^2B_{r,i+\half,j,k}-r_{i-\half} ^2B_{r,i-\half,j,k}}{r_i^2\Delta r}\nonumber\\
&+\frac{\sin\theta_{j+\half}B_{\theta,i,j+\half,k}-\sin\theta_{j-\half}B_{\theta,i,j-\half, k}}{r_i\Delta\cos\theta}+\frac{B_{\phi,i,j,k+\half}-B_{\phi,i,j,k-\half}}{r_i\sin\theta_j\Delta\phi}.\label{ef22}
\end{align}
Next we need to find $\Phi_{i,j,k}$ that satisfies the discrete Laplace equation Equation \eqref{ef22} and boundary conditions. If it is substituted into $\Phi=0$, a non-zero residual $R_{i,j,k}$ will be generated due to the non-uniformity of inner boundary condition. Construct a new boundary value problem
\[(\nabla^2\Phi)_{i,j,k}=R_{i,j,k},\]
\[\Phi_{0,j,k}=\Phi_{1,j,k},\]
and solve it by iterative method, then the solution and the initial boundary conditions just constitute the required potential field solution.
\subsubsection{\texttt{pfsspy} algorithm and construction of eigenvalue problem in finite difference form \cite{Pfsspy}}
For spherical coordinates $(r, \theta, \phi)$, note $\rho=\ln(r)$, $s=\cos\theta$. The algorithm uses a grid composed of equidistantly divided $\rho$, $s$, $\phi$ for calculation, and the Lamé coefficient of the orthogonal curvilinear coordinate system $|\mathrm{d}\boldsymbol r/\mathrm{ d}\rho|$, $|\mathrm{d}\boldsymbol r/\mathrm{d}s|$, $|\mathrm{d}\boldsymbol r/\mathrm{d}\phi|$ are respectively
\[h_\rho = r = \mathrm{e}^\rho,\quad h_s = \frac{r}{\sin\theta} = \frac{\mathrm{e}^\rho}{\sqrt{1-s ^2}}, \quad h_\phi = r\sin\theta = \mathrm{e}^\rho\sqrt{1-s^2}.\]
The general strategy for calculating the magnetic field here is to construct its vector potential according to the passive property, i.e. to assume
\[\Bb=\nabla\times\Ab',\quad \Ab' = \nabla\times\big(\psi' \evr\big),\]
then in curvilinear coordinates
\begin{equation}
\Bb= -\Delta_\perp\psi'\evr + \frac{1}{h_\phi}\dy_\phi\left(\frac{1}{h_\rho}\dy_\rho\psi'\right) \evp + \frac{1}{h_s}\dy_s\left(\frac{1}{h_\rho}\dy_\rho\psi'\right)\evs.
\label{e1}
\end{equation}
And from the irrotational property, $\Bb$ can be expressed as gradient of a scalar. Considering Equation \eqref{e1}, it can be first noted as
\[\Bb = \nabla\big(\frac1{h_\rho}\dy_\rho\psi'\big) + f'(\rho),\]
where $f'(\rho)$ is a unary function of $\rho$.
Let $\psi=\psi' + h_\rho\int f'(\rho)\mathrm{d}\rho$, $\Ab = \nabla\times(\psi \evr)$, then
\begin{equation}
\Bb = \nabla\big(\frac1{h_\rho}\dy_\rho\psi\big). \label{e2}
\end{equation}
It is easy to verify that there's also
\begin{equation}
\Bb=\nabla\times\Ab=-\Delta_\perp\psi\evr + \frac{1}{h_\phi}\dy_\phi\left(\frac{1}{h_\rho}\dy_\rho\psi\right)\evp + \frac{1}{h_s}\dy_s\left(\frac{1}{h_\rho}\dy_\rho\psi\right)\evs.
\label{e3}
\end{equation}
Then by Equation \eqref{e2} and \eqref{e3},
\begin{equation}
\nabla^2_\perp\psi = -\frac{1}{h_\rho}\dy_\rho\left(\frac{1}{h_\rho}\dy_\rho\psi\right).
\label{eqn:psi}
\end{equation}
Next we only need to solve $\psi$ according to Equation \eqref{eqn:psi} and then calculate $\Ab$ and $\Bb$.
Note the grid point as $(\rho^k, s^j, \phi^i)$ and define the edge lengths $L_\rho^{k+\half,j,i}$, $L_s^{k,j+\half,i}$, $L_\phi^{k,j,i+\half}$ then Equation \eqref{eqn:psi} can be discretized as
\begin{align}
U^{j+\half}\big(\psi^{k,j+\half,i+\thr} - \psi^{k,j+\half,i-\half} \big) + V^{j+1 }\psi^{k,j+\thr,i+\half} + V^{j}\psi^{k,j-\half,i+\half}\nonumber\\
 - \Big(2U^{j+\half} + V^{j+1} + V^{j}\Big)\psi^{k,j+\half,i+\half}\nonumber\\
 = -\frac{c(\dr)\ex^{2\rho^k}}{L_\rho^{k,j+\half,i+\half}}\left(\frac{\psi^{k+ 1,j+\half,i+\half} - \psi^{k,j+\half,i+\half}}{L_\rho^{k+\half,j+\half,i+\half}} - \frac{\psi^{k,j+\half,i+\half} - \psi^{k-1,j+\half,i+\half}}{L_\rho^{k-\half,j+\half,i+\half} } \right),
 \label{eqn:main}
\end{align}
where
\[U^{j+\half} = \left(\frac{L_s}{\ds\dph L_\phi}\right)^{j+\half}, V^j = \left(\frac{L_\phi }{\ds\dph L_s}\right)^j, c(\dr) = \frac{2\ex^{\dr/2}}{\ex^{\dr} + 1} = \mathrm{sech}\left(\frac{\dr}{2}\right).\]
Boundary conditions in \texttt{pfsspy} algorithm are also set by a layer of ghost cells. Next it‘s only necessary to solve the $\nr\ns\nph\times\nr\ns\nph$ order linear equations composed of Equation \eqref{eqn:main}. Suppose there are eigenfunctions of the form
\begin{equation}
\psi^{k,j+\half,i+\half} = f^kQ_{lm}^{j+\half}\,\ex^{2\pi I mi/\nph}.
\label{eqn:eig}
\end{equation}
where $k$ in $f^k$ represents a power, $I$ is imaginary unit, and $Q$ is a set of standard orthogonal functions about $l$.
Substitute Equation \eqref{eqn:eig} into \eqref{eqn:main} to get the tridiagonal eigenvalue problem
\begin{equation}
-V^{j}Q^{j-\half}_{lm} + \left(V^{j} + V^{j+1}+ 4U^{j+\half}\sin^2\left( \tfrac{\pi m}{\nph}\right) \right)Q^{j+\half}_{lm} - V^{j+1}Q^{j+\thr}_{lm} = \lambda_ {lm}Q_{lm}^{j+\half},
\end{equation}
thus $f$ can be obtained from $\lambda_{lm}$ by solving the quadratic equation
\begin{equation}
\lambda_{lm} = \frac{c(\dr)}{\mathrm{e}^{\dr/2} - \mathrm{e}^{-\dr/2}} \left(\frac{f -1}{\mathrm{e}^{\dr} - 1}-\frac{1-f^{-1}}{1-\mathrm{e}^{-\dr}}\right).
\end{equation}
For each $l$, $m$, the two solutions can be expressed as $f_{lm}^+, f_{lm}^-$.
The potential field $\psi$ can be written as a linear combination of these two sets of radial eigenfunctions:
\begin{equation}
\psi^{k,j+\half,i+\half} = \sum_{l=0}^{\ns-1}\sum_{m=0}^{\nph-1}\Big[c_{lm} (f_{lm}^+)^k + d_{lm}(f_{lm}^-)^k) \Big] Q_{lm}^{j+\half}\ex^{2\pi I mi/\nph},
\label{eqn:psisum}
\end{equation}
where coefficients $c_{lm}$ and $d_{lm}$ are determined by boundary conditions.
In fact, with the improvement of grid density, $Q_{lm}^{j+\half}$ as a function of $\theta$ should converge to the corresponding associated Legendre polynomial $P_l^m(\cos\theta)$ in the sense of normalization.
\subsection{Alfvén Wave Solar Model \cite{Van}}
We use Alfvén Wave Solar Model (AWSoM) based on fundamental equations of magnetohydrodynamics as a reference that considers as many physical processes and different particle motion properties in solar-terrestrial space as possible. This is a global model from upper chromosphere to  heliosphere, dealing with coronal heating and solar wind acceleration with Alfvén wave turbulence, making Poynting flux proportional to magnetic field by injecting Alfvén wave energy at the inner boundary. The model also uses photospheric measurements to simulate the 3D magnetic field topology, but does not impose a boundary between open and closed field lines.

\begin{figure}[htbp]
\begin{center}
\includegraphics[width=0.5\textwidth]{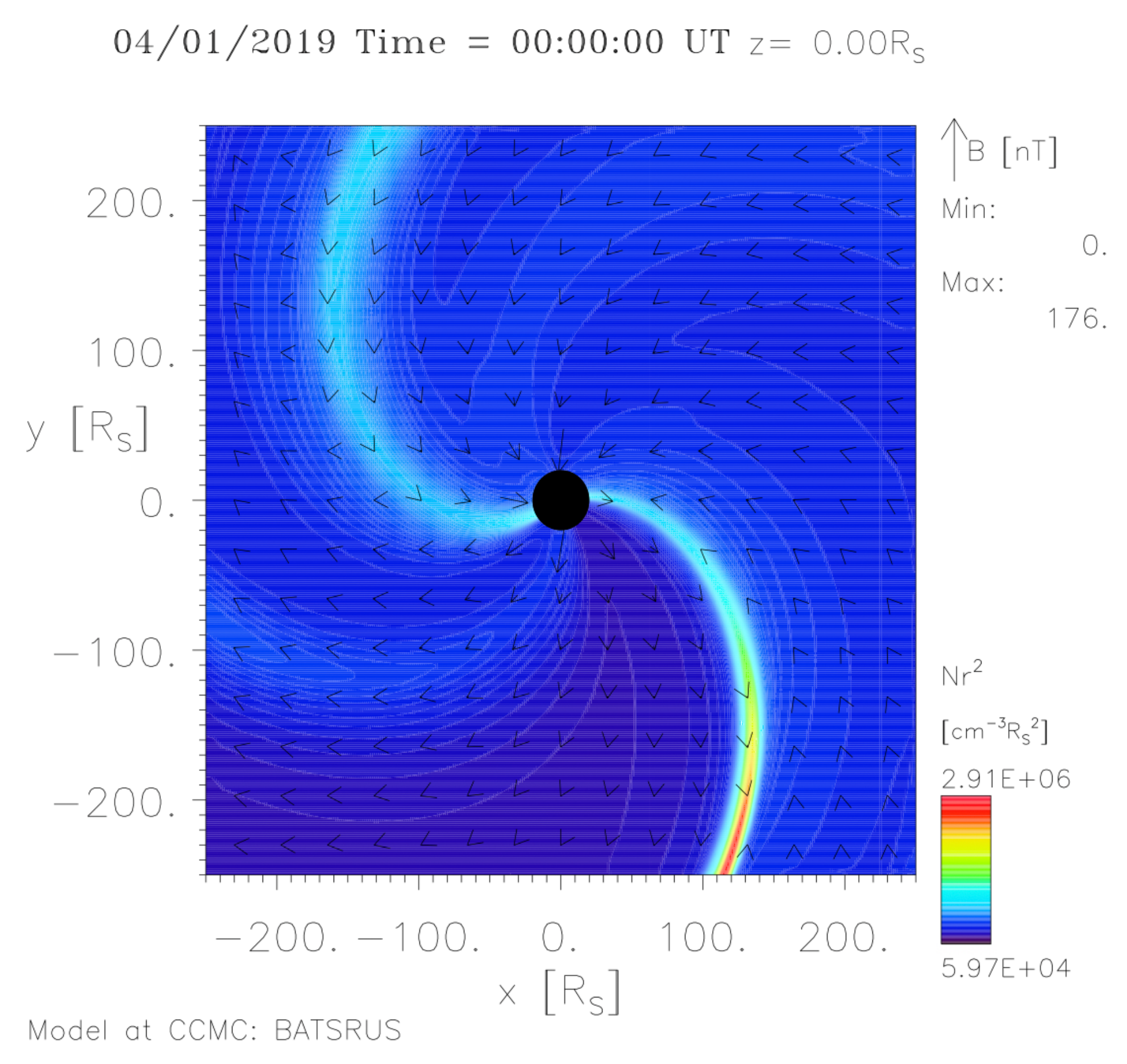}
\caption{Density and magnetic field distribution on solar equatorial plane deduced by AWSoM.}
\label{mhd2215_16_equator}
\end{center}
\end{figure}

\begin{figure}[htbp]
\begin{center}
\includegraphics[width=0.5\textwidth]{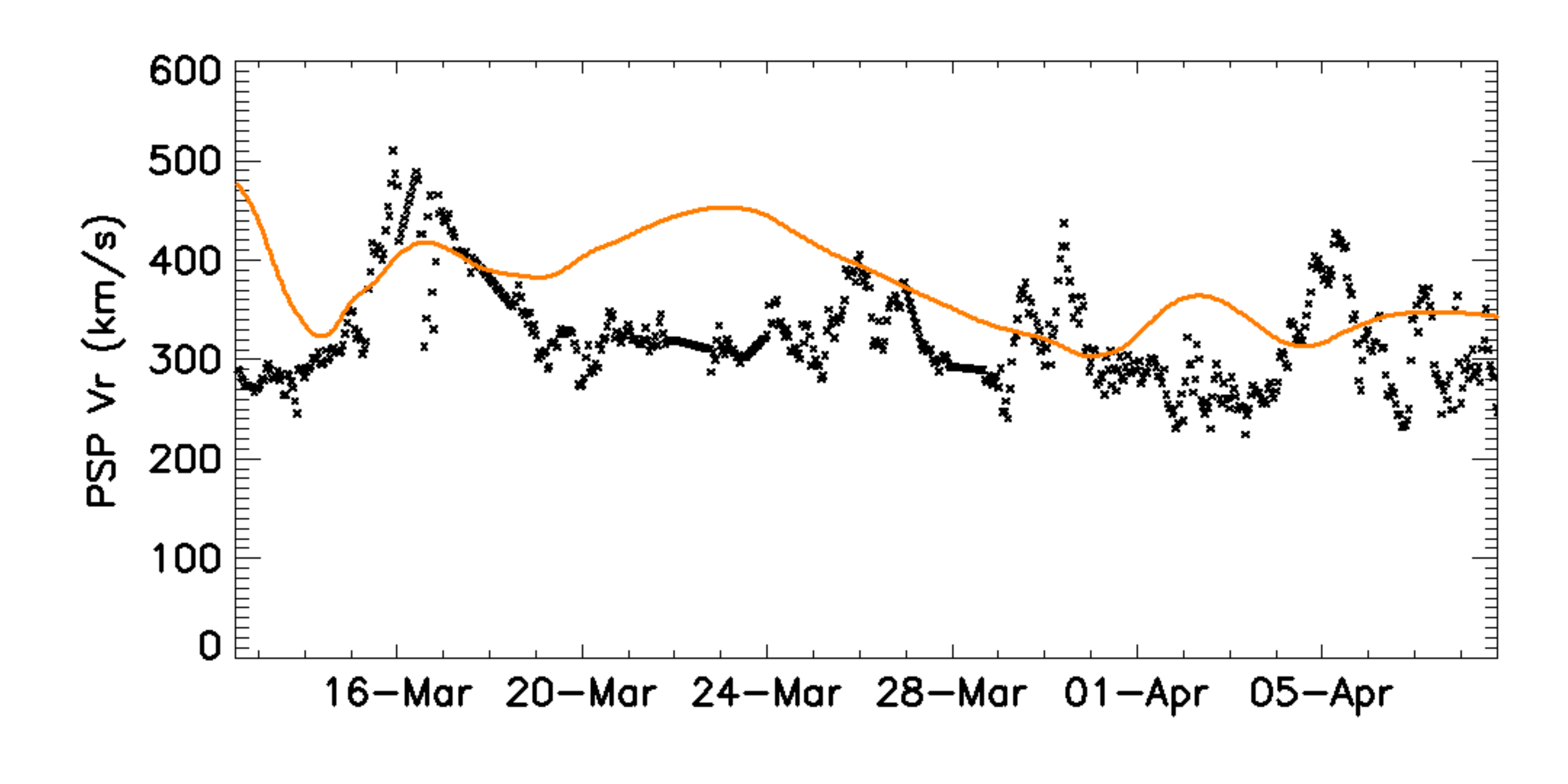}
\caption{Radial solar wind speed measured by PSP (black) and deduced by AWSoM (orange). }
\label{vr}
\end{center}
\end{figure}

Figure \ref{mhd2215_16_equator} shows the deduced density and magnetic field distribution on the solar equatorial plane in CR2215. Figure \ref{vr} depicts the solar wind velocity calculated by this MHD model against corresponding PSP observations during CR2215, where their mean values are 373.2 km/s and 322.7 km/s respectively, with a root mean square error of 79.1 km/s.

\section{Evaluation of Coronal and Interplanetary Magnetic Field Modeling}
\begin{figure}[htbp]
\begin{center}
\includegraphics[width=0.5\textwidth]{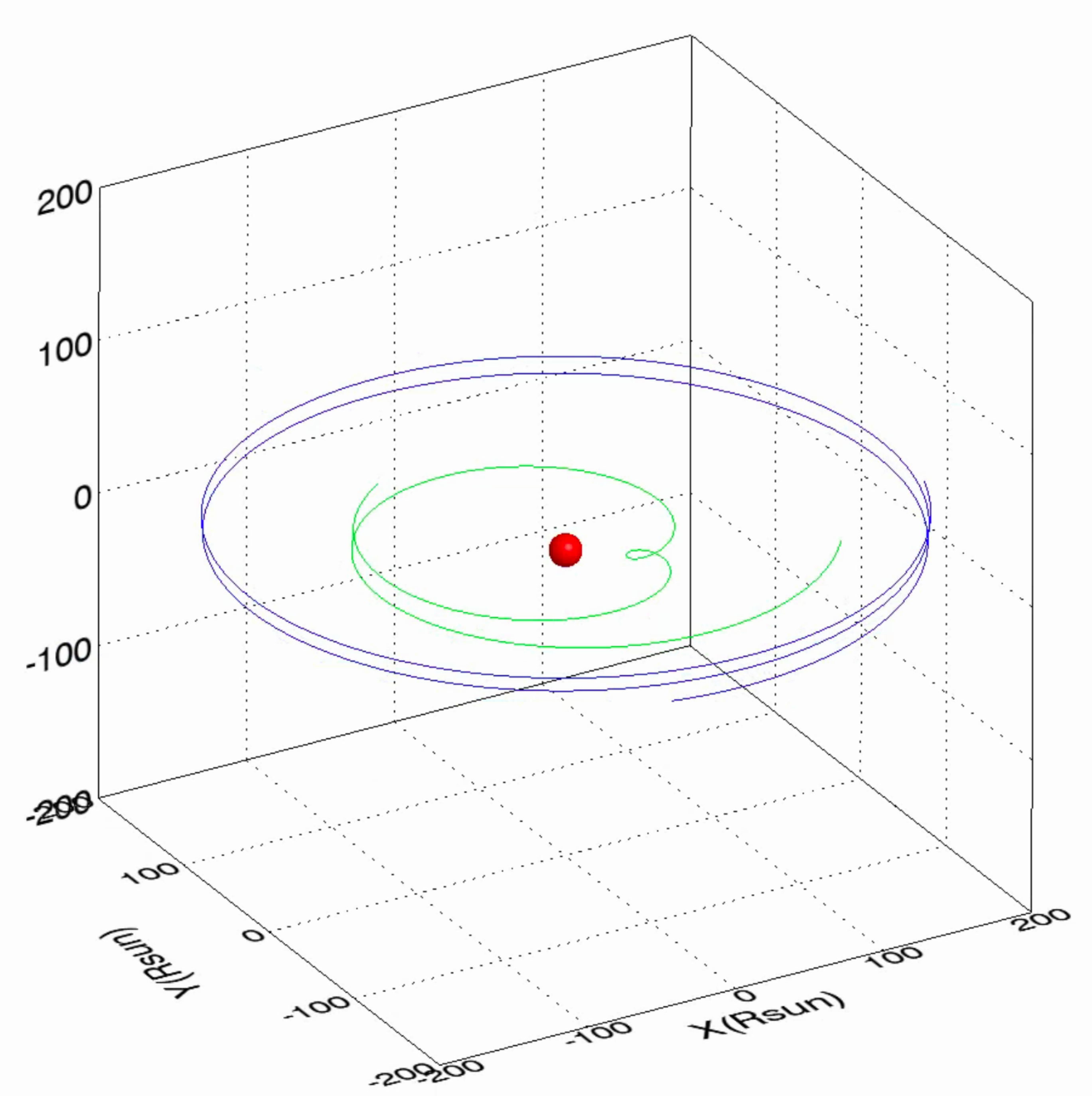}
\caption{The trajectory of PSP (green) and Earth (purple) from 2018 Oct. 1 to Nov. 30.}
\label{fig:psp_orbit_E1}
\end{center}
\end{figure}

In Part 4.1, we calculated the magnetic field results of different models based on the observation of CR2210 which is around the first perihelion of PSP (2018 November 5), and studied the influence of relevant parameters and synoptic maps. The performance of some models at 1AU is briefly discussed in Part 4.2. Part 4.3 analyzes the importance of PSP's near-solar detection. Then in Part 4.4 we made some attempts to improve the PFSS model and applied the method to CR2215 around the second perihelion of PSP (2019 April 4). PSP had similar paths in the two periods of interest, with their closest approaches to the Sun around 0.167AU. Figure \ref{fig:psp_orbit_E1} shows the trajectory of PSP and Earth in carrington system around the first one.

\begin{figure}[htbp]
\begin{center}
\hbox{
\includegraphics[width=0.5\textwidth,clip=]{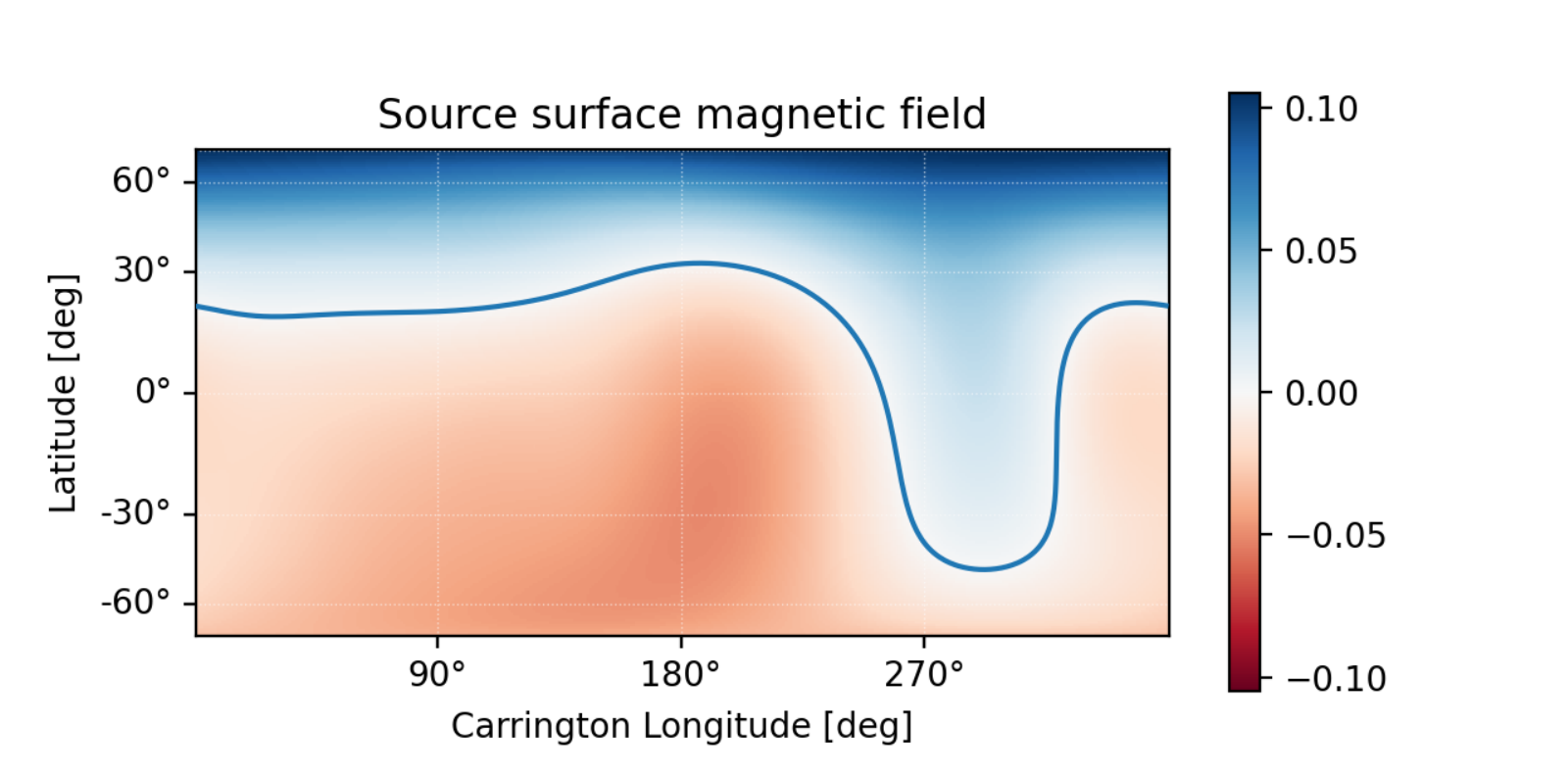}
\includegraphics[width=0.5\textwidth,clip=]{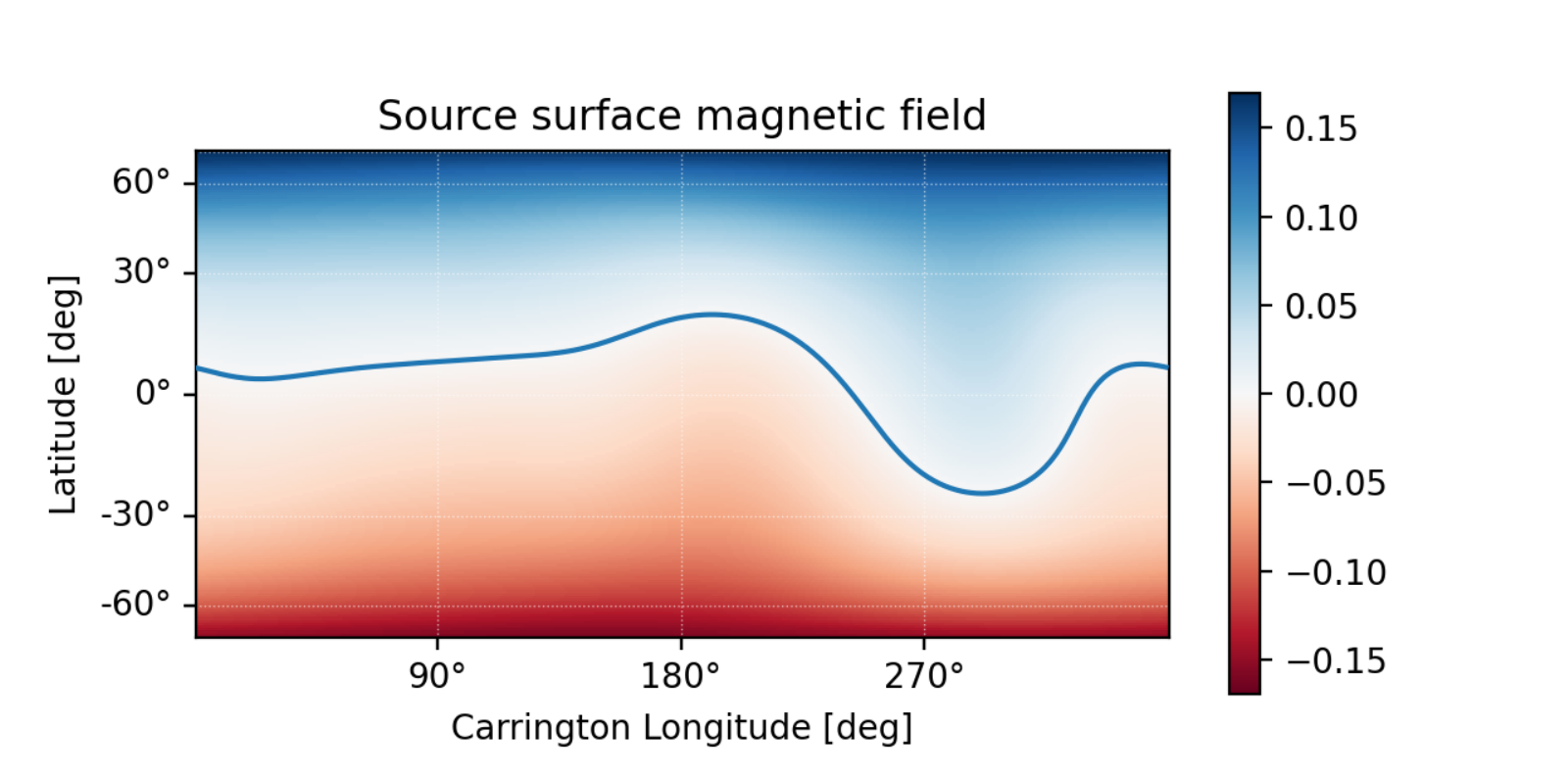}
}
\caption{The polarity distribution of the source surface magnetic field obtained from GONG standard magnetogram (left) and zero-point corrected magnetogram (right).}
\label{fig:gong_comp}
\end{center}
\end{figure}

Regarding the data selection of GONG, Figure \ref{fig:gong_comp} compares the magnetic polarity distribution of source surface when the potential field is calculated using the standard  magnetogram and the zero-point corrected magnetogram in the same period. It can be seen that it is necessary to correct the zero-point uncertainty. Therefore all subsequent simulations will use the zero-corrected version when inputting the GONG photosphere magnetogram.

The simulated interplanetary magnetic field is usually found to be underestimated when compared with in-situ observation, i.e. ``open flux problem'' \cite{Linker}. Because the magnetic flux originating from solar polar region may have a non-negligible contribution to the interplanetary magnetic field, but due to the limitation of observation conditions, this part of data is often missing or inaccurate, which may lead to a low result when using the photosphere magnetic map for extrapolation. This problem exists in many models and has not been solved, but the effect can be partially corrected either by adding a polar field to the model or by multiplying the simulation results by an appropriate coefficient \cite{Linker,Riley}. For convenience of comparison, simulation curves in this paper are scaled up to have the same average absolute value of $B_r$ as PSP observed data.

\subsection{Modeling and Parameter Analysis Based on the First Perihelion Observation of PSP}
In this section we use 4 algorithms introduced in Section 3 to calculate the coronal and interplanetary magnetic fields, and compare them with in-situ observation of PSP. The time range of CR2210 is from 2018 October 26 20:53:35 to November 23 04:13:09.

\begin{figure}[htbp]
\begin{center}
\includegraphics[width=0.5\textwidth]{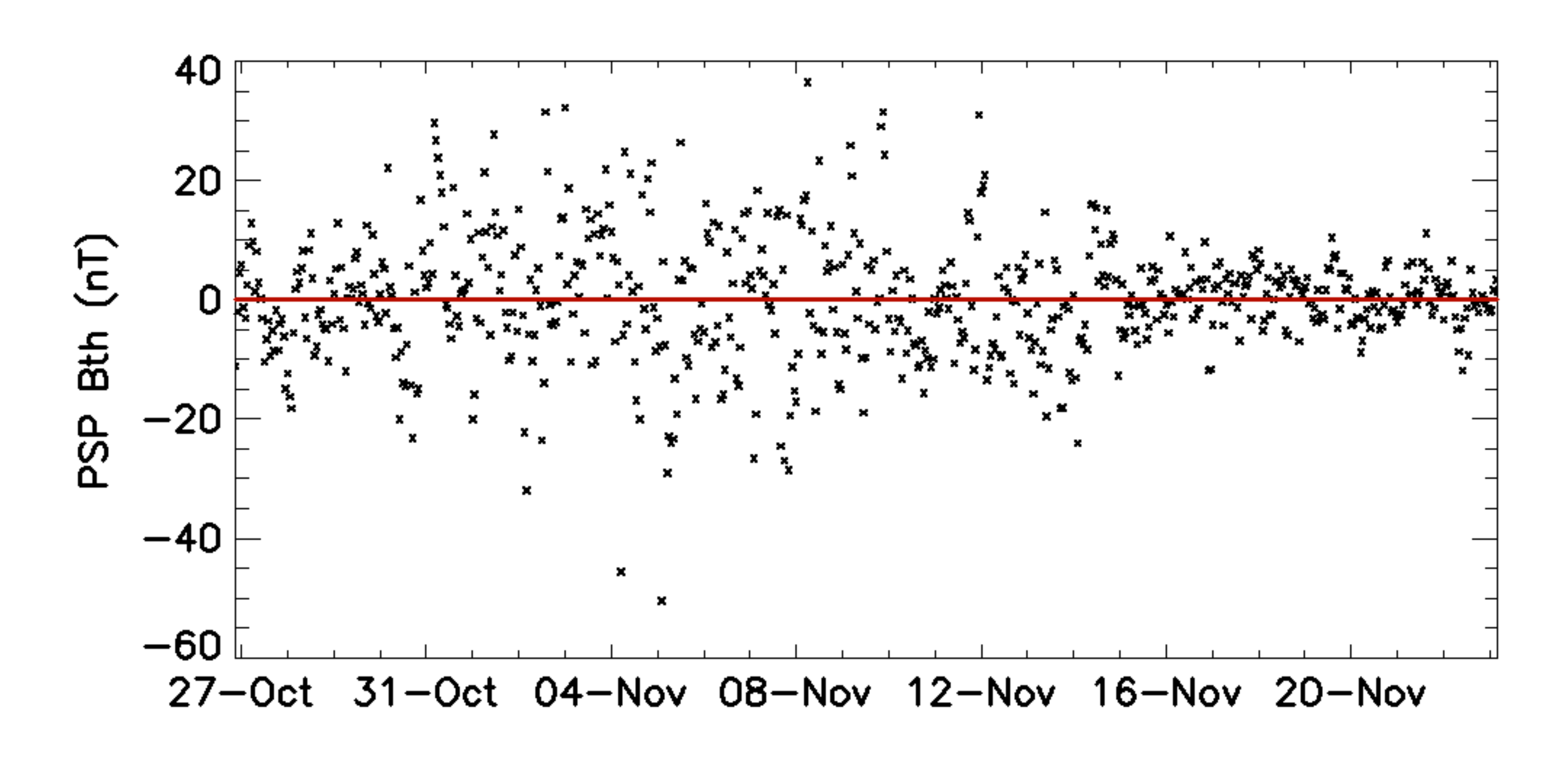}
\caption{The extrapolated $B_\theta$ in PFSS model (red) and PSP in-situ data (black). }
\label{pfss_bth}
\end{center}
\end{figure}

As shown in Figure \ref{pfss_bth}, the $B_\theta$ component of interplanetary magnetic field obtained by PFSS model is always zero, which has a RMSE (root mean square error) of about 10.1777 nT compared with the observed data of PSP in CR2210. For brevity we will avoid displaying it repeatedly in this part.

\subsubsection{Spherical harmonic method results}
\begin{table}[htb]
\centering
\caption{LINFF experiment records. From left to right are successively the number of truncated terms, the number of grids in three directions, the position of source surface (referring to solar radius), magnetogram, the scaling coefficient, the polarity coincidence rate of $B_r$ and $B_\phi$ components with observed data, and the RMSE of scaled $B_r$, $B_\phi$ and $B$ relative to observed data.}
\label{tab:1}
\begin{tabular}{ccccccccccccc}
\toprule
\#&l&$N_r$&$N_\theta$&$N_\phi$&Rss&Input&Scale&$P_r$&$P_\phi$&RMSE($B_r$)&RMSE($B_\phi$)&RMSE($B$)\\
\midrule
  1&24&  50&  90&180&2.0&    HMI&3.25958&0.884146&0.762195&14.7592&12.7902&11.5167\\
  2&24&  75&  90&180&2.5&    HMI&8.76151&0.893293&0.771341&19.4168&13.1917&18.1216\\
  3&24&150&180&360&2.5&    HMI&8.75295&0.893293&0.771341&19.4094&13.1918&18.1122\\
  4&24&100&180&360&2.0&    HMI&3.25737&0.884146&0.762195&14.7597&12.7910&11.5179\\
  5&12&100&180&360&2.0&    HMI&3.25702&0.884146&0.762195&14.7598&12.7911&11.5177\\
  6&12&150&180&360&2.5&    HMI&8.75284&0.893293&0.771341&19.4090&13.1917&18.1117\\
  7&12&  75&  90&180&2.5&    HMI&8.76141&0.893293&0.771341&19.4165&13.1917&18.1212\\
  8&12&  50&  90&180&2.0&    HMI&3.25924&0.884146&0.762195&14.7592&12.7903&11.5165\\
  9&12&  50&  90&180&2.0&GONG&3.68910&0.873476&0.754573&14.7237&12.7941&11.4225\\
10&12&  75&  90&180&2.5&GONG&10.8356&0.888719&0.772866&20.2375&13.2663&19.1685\\
11&12&150&180&360&2.5&GONG&10.8233&0.888719&0.772866&20.2267&13.2661&19.1552\\
12&12&100&180&360&2.0&GONG&3.68647&0.873476&0.754573&14.7244&12.7950&11.4239\\
13&24&100&180&360&2.0&GONG&3.68674&0.873476&0.754573&14.7239&12.7949&11.4236\\
14&24&150&180&360&2.5&GONG&10.8234&0.888719&0.772866&20.2269&13.2661&19.1554\\
15&24&  75&  90&180&2.5&GONG&10.8356&0.888719&0.772866&20.2377&13.2663&19.1688\\
16&24&  50&  90&180&2.0&GONG&3.68937&0.873476&0.754573&14.7233&12.7940&11.4222\\
\bottomrule
\end{tabular}
\end{table}

\begin{figure}[htbp]
\begin{center}
\includegraphics[width=1\textwidth]{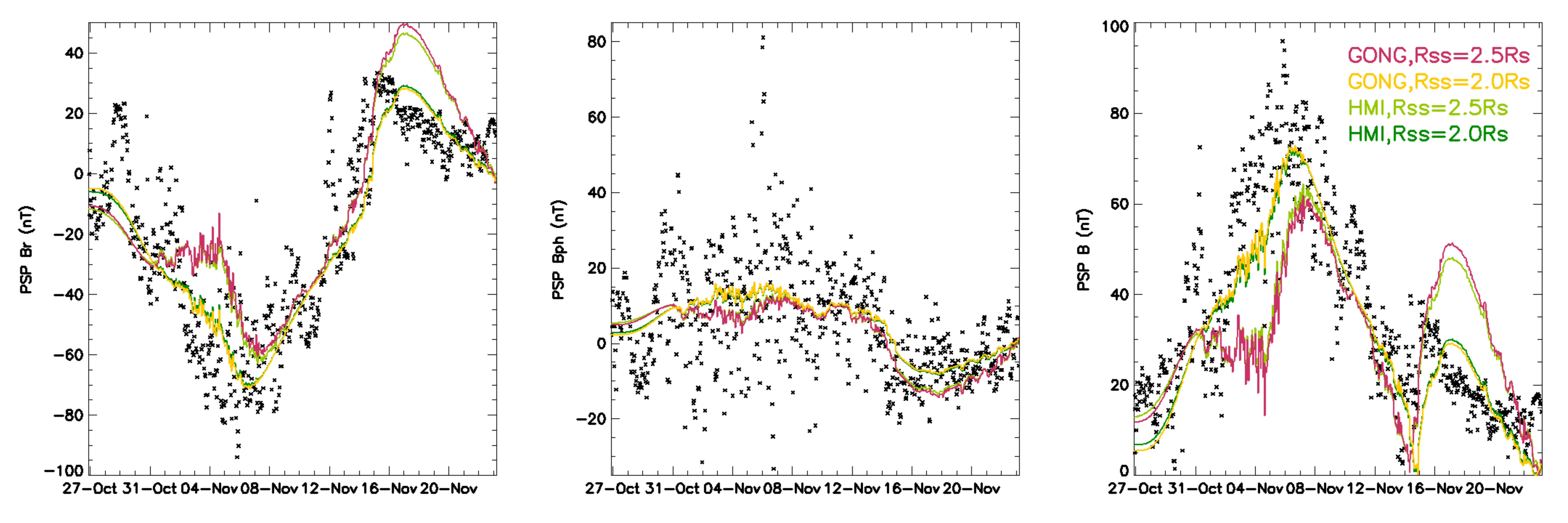}
\caption{The spherical harmonic function extrapolated magnetic field compared with PSP in-situ data (black). }
\label{c1}
\end{center}
\end{figure}

\begin{figure}[htbp]
\begin{center}
\vbox{
\includegraphics[width=1\textwidth]{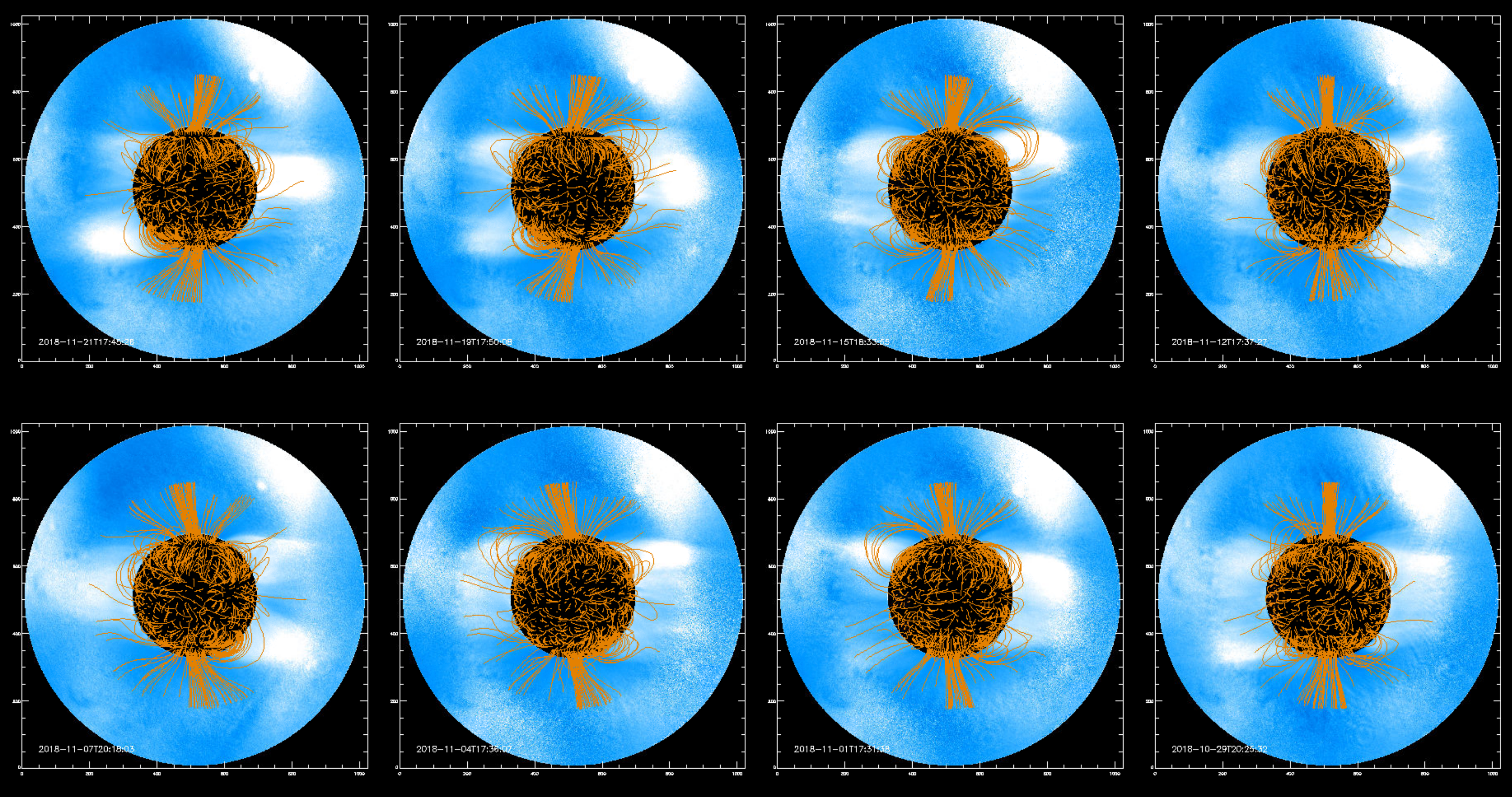}
\includegraphics[width=1\textwidth]{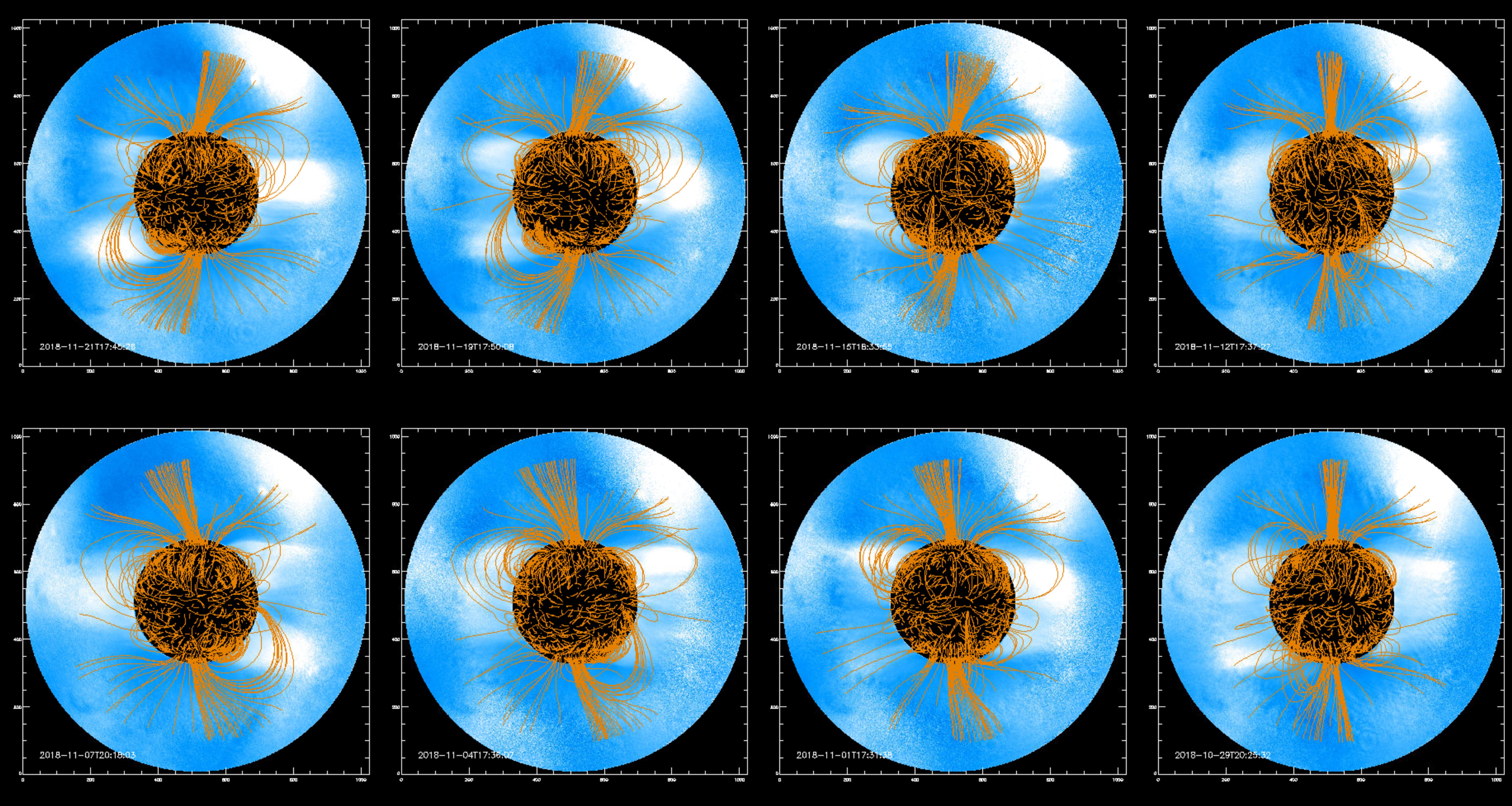}}
\caption{The coronal magnetic field structure at different time in CR2210 with a designated source surface at 2.0~Rs (upper two rows) and 2.5~Rs (lower two rows) superimposed on K-COR white light images.}
\label{c29c30}
\end{center}
\end{figure}

As for analytical method of potential field model, the standard potential field solution tool in LINFF code developed by Thomas Wiegelmann is used here. Since this algorithm requires quite a lot of storage space, only some short truncations could be analyzed. Figure \ref{c1} shows a comparison of these results with observed magnetic field provided by PSP. Parameters are shown in Table \ref{tab:1}.The number of truncated terms and mesh density have very slight effects on the results in the range of values we selected. As we can see in Figure \ref{c1}, the curves for same magnetogram and same source surface position almost overlap. HMI magnetogram with smaller scaling coefficients seem to have better performance in the reconstruction of magnetic field strength. Each simulation reliably reproduced magnetic polarity, with the results at 2.5 Rs slightly better than those at 2.0Rs but they were fairly close. Figure \ref{c29c30} shows field lines of the 13th and 14th simulations against K-COR white light images from several angles, where open field lines versus coronal holes, closed field lines versus streamers are basically corresponding. However, in the reconstruction of magnetic field intensity and variation, the model performance is significantly better when the source surface is set at 2.0 Rs than at 2.5 Rs.

\subsubsection{Finite difference iterative method results}
\begin{table}[htb]
\centering
\caption{FDIPS experiment records. From left to right are successively the number of grids in three directions, the position of source surface (referring to solar radius), magnetogram, the scaling coefficient, the polarity coincidence rate of $B_r$ and $B_\phi$ components with observed data, and the RMSE of scaled $B_r$, $B_\phi$ and $B$ relative to observed data.}
\label{tab:fd2}
\begin{tabular}{cccccccccccc}
\toprule
\#&$N_r$&$N_\theta$&$N_\phi$&Rss&Input&Scale&$P_r$&$P_\phi$&RMSE($B_r$)&RMSE($B_\phi$)&RMSE($B$)\\
\midrule
1&150&180&360&2.5&GONG&14.9516&0.899390&0.771341&18.6481&13.1826&18.2914\\
2&100&180&360&2.0&GONG&5.84323&0.887195&0.759146&14.7839&12.8209&11.8553\\
3&150&180&360&2.5&    HMI&13.3404&0.893293&0.771341&19.4599&13.2055&18.3009\\
4&150&180&360&2.0&    HMI&5.19182&0.882622&0.760671&15.1098&12.8271&11.8508\\
\bottomrule
\end{tabular}
\end{table}

\begin{figure}[htbp]
\begin{center}
\includegraphics[width=1\textwidth]{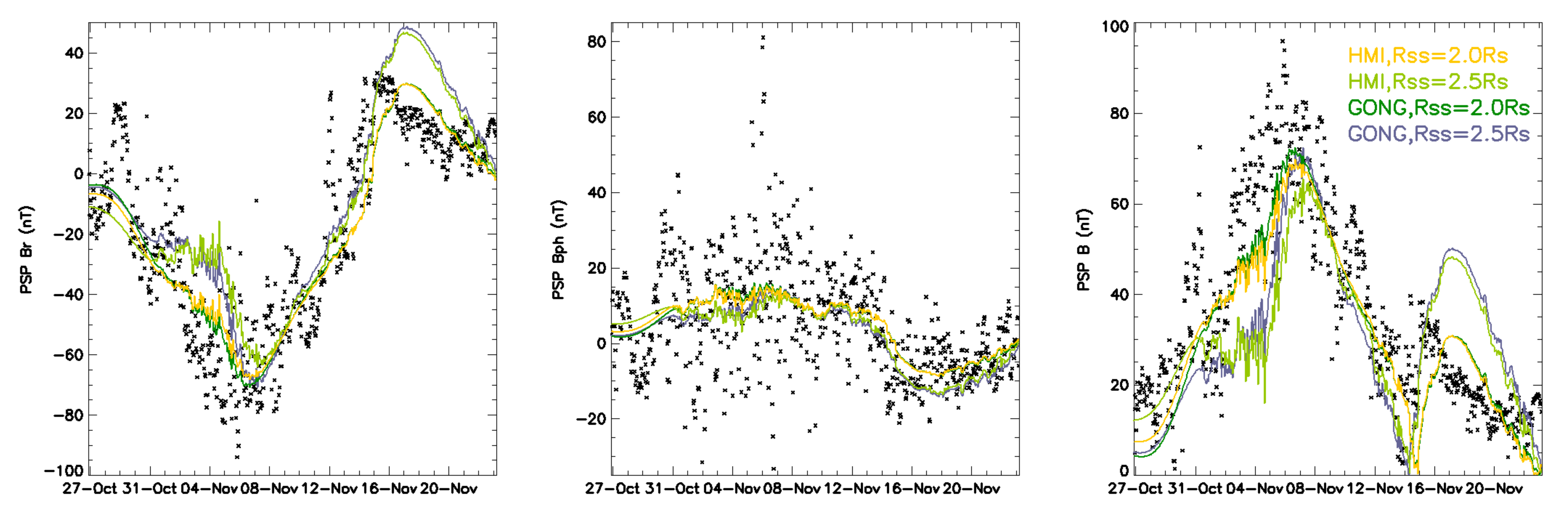}
\caption{Comparison of FDIPS magnetic field simulation and PSP in-situ data (black).}
\label{fd1}
\end{center}
\end{figure}

The results of FDIPS are shown in Figure \ref{fd1}. The iteration accuracy is to relative error of $10^{-10}$ and other input parameters are shown in Table \ref{tab:fd2}. The influence of magnetogram and source surface position setting on the model is basically the same as that in LINFF algorithm, but it can be inferred that this method is slightly deficient about reconstruction of magnetic field intensity, because the scaling coefficients are relatively high and dwindling the relative error of iterative calculation to $10^{-15}$ or increasing the number of radial grid points to 1000 cannot further reduce them. Moreover, if the mesh density in these four simulations is reduced by half, the magnetic polarity changes will be completely unreliable. Considering that grid density of model is limited by the spatial resolution of the photosphere magnetic field measurement, it is inevitable that direct difference method like this will omit more details in observed data.

\subsubsection{Finite difference eigenvalue method results}
\begin{table}[htb]
\centering
\caption{pfsspy experiment records. From left to right are successively the number of grids in three directions, the position of source surface (referring to solar radius), magnetogram, the scaling coefficient, the polarity coincidence rate of $B_r$ and $B_\phi$ components with observed data, and the RMSE of scaled $B_r$, $B_\phi$ and $B$ relative to observed data.}
\label{tab:2}
\begin{tabular}{cccccccccccc}
\toprule
\#&$N_r$&$N_\theta$&$N_\phi$&Rss&Input&Scale&$P_r$&$P_\phi$&RMSE($B_r$)&RMSE($B_\phi$)&RMSE($B$)\\
\midrule
1&150&180&360&2.5& HMI&8.39061&0.896341&0.771341&17.1127&12.9424&15.4212\\
2&100&180&360&2.0& HMI&3.55109&0.891768&0.763719&14.8055&12.7654&11.8749\\
3&100&180&360&2.0&GONG&3.99026&0.884146&0.765244&14.8394&12.7594&11.4648\\
4&150&180&360&2.5&GONG&9.66459&0.891768&0.772866&17.4822&12.9596&15.3989\\
5& 75& 90&180&2.5&GONG&9.53747&0.893293&0.771341&18.2963&13.0382&16.3696\\
6& 50& 90&180&2.0&GONG&3.90028&0.884146&0.765244&14.9576&12.7592&11.6256\\
7&100&360&720&2.0&    HMI&3.44005&0.885671&0.763719&14.9135&12.7766&11.3967\\
\bottomrule
\end{tabular}
\end{table}

\begin{figure}[htbp]
\begin{center}
\includegraphics[width=1\textwidth]{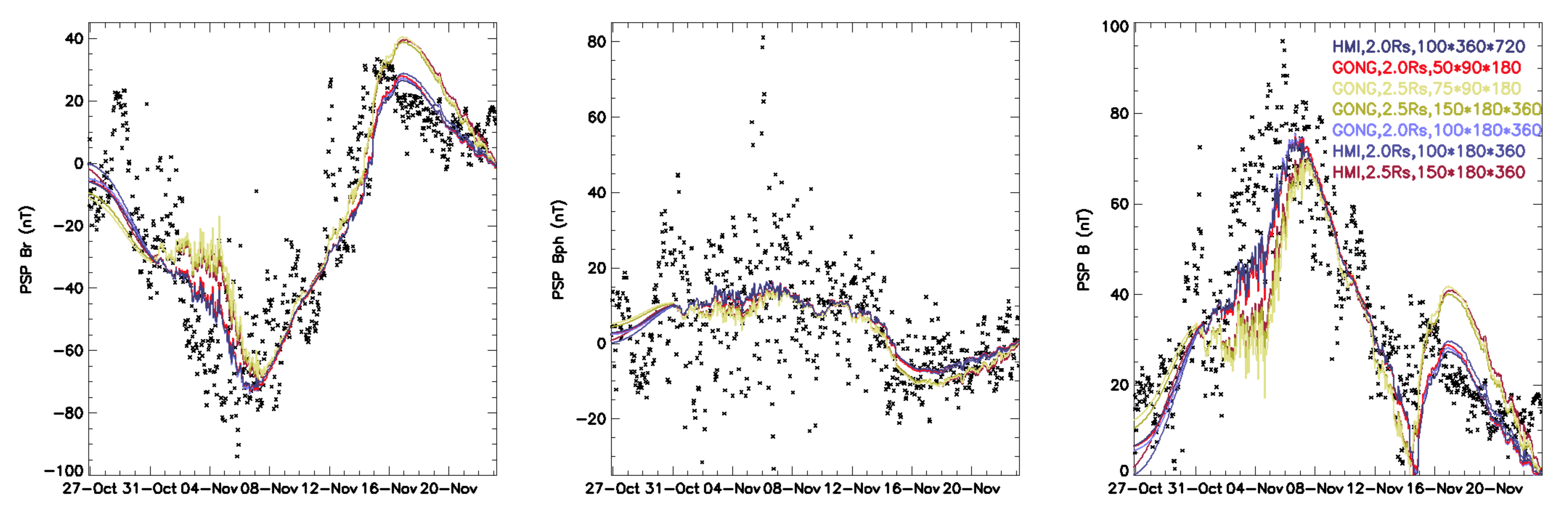}
\caption{Comparison of pfsspy magnetic field simulation and PSP in-situ data (black).}
\label{c2}
\end{center}
\end{figure}

The results obtained by pfsspy algorithm are shown in Figure \ref{c2} and Table \ref{tab:2} is for detailed input parameters. The effect of source surface position and magnetogram setting is similar to that in the previous two algorithms. This time, GONG magnetogram still yielded usable results when extrapolating with a lower mesh density (the 5th and 6th in Table \ref{tab:2}) while HMI didn't. It is also worth noting that pfsspy gives magnetic field intensity higher thus also closer to observed data than FDIPS does. The higher mesh density in the 7th simulation further optimizes its performance in field strength reconstruction. It can be seen that a tactical separation of variables according to the structure of analytical solution is very helpful to improve the finite difference method.

Although it is possible to obtain better results by refining the grid based on HMI synoptic maps with high spatial resolution, this operation requires the support of huge computing resources. Consistent with the complexity of the calculation mechanism, LINFF, FDIPS and pfsspy consume about several hours to more than ten hours, tens of minutes and tens of seconds for a calculation respectively. And under the usual conditions, a little attempt to refine the grid is only feasible in pfsspy algorithm.

\subsubsection{MHD results}
\begin{figure}[htbp]
\begin{center}
\includegraphics[width=1\textwidth]{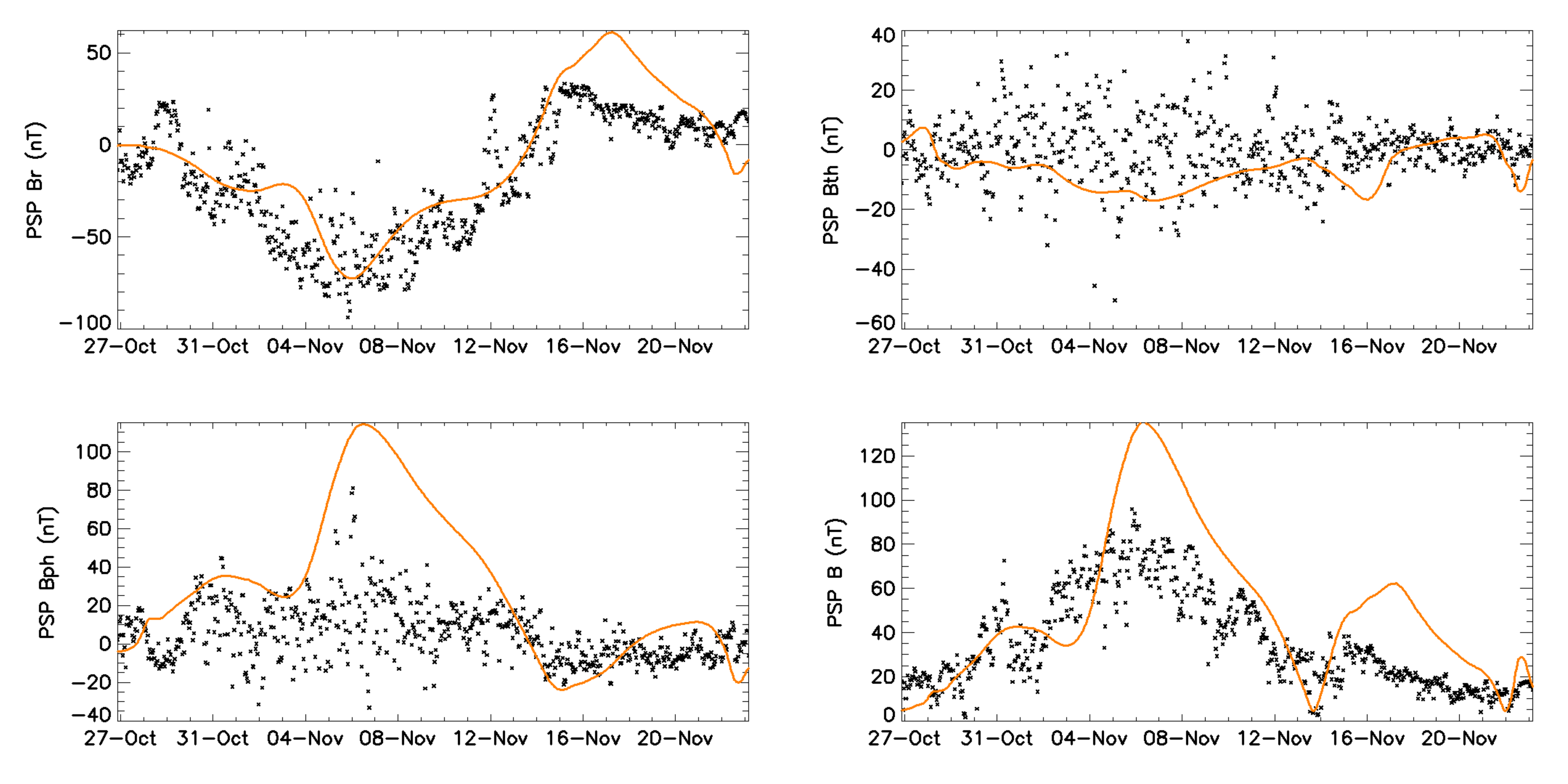}
\caption{Comparison of AWSoM simulation (orange) and PSP in-situ data (black).}
\label{c4}
\end{center}
\end{figure}

\begin{table}[htb]
\centering
\caption{AWSoM experiment records. From left to right are successively the scaling coefficient, the polarity coincidence rate of three components with observed data, and the RMSE of scaled $B_r$, $B_\theta$, $B_\phi$ and $B$ relative to observed data.}
\label{tab:mhd}
\begin{tabular}{cccccccc}
\toprule
Scale & $P_r$ & $P_\theta$ & $P_\phi$ & RMSE($B_r$) & RMSE($B_\theta$) & RMSE($B_\phi$) & RMSE($B$)\\
\midrule
4.40752&0.867378&0.503049&0.643293&21.1803&13.8542&41.0425&26.3491\\
\bottomrule
\end{tabular}
\end{table}

The main input parameters we use in AWSoM MHD model are Poynting ratio$=0.7\times 10^6$ J/(m$^2$sT), which is ratio of Poynting flux to magnetic field strength at the photosphere level, and Coronal Heating$=1.5\times 10^5$ mT$^{1/2}$, which is perpendicular correlation length times the square root of local magnetic field intensity. They are assumed to be constants in this model and selected according to developers' recommendation which is also experienced almost the optimal selection. The inner boundary is at 1.1 Rs. Simulation results are shown in Figure \ref{c4} and Table \ref{tab:mhd}. It's clear that previous PFSS simulations using solar wind velocity observation show much more accurate details. Although MHD model provides more information than PFSS about $B_\theta$, which is actually of the smallest magnitude, it seems difficult to give a sufficiently reliable prediction.

\subsection{Results at 1 AU}
\begin{figure}[htbp]
\begin{center}
\vbox{
\includegraphics[width=1\textwidth,clip=]{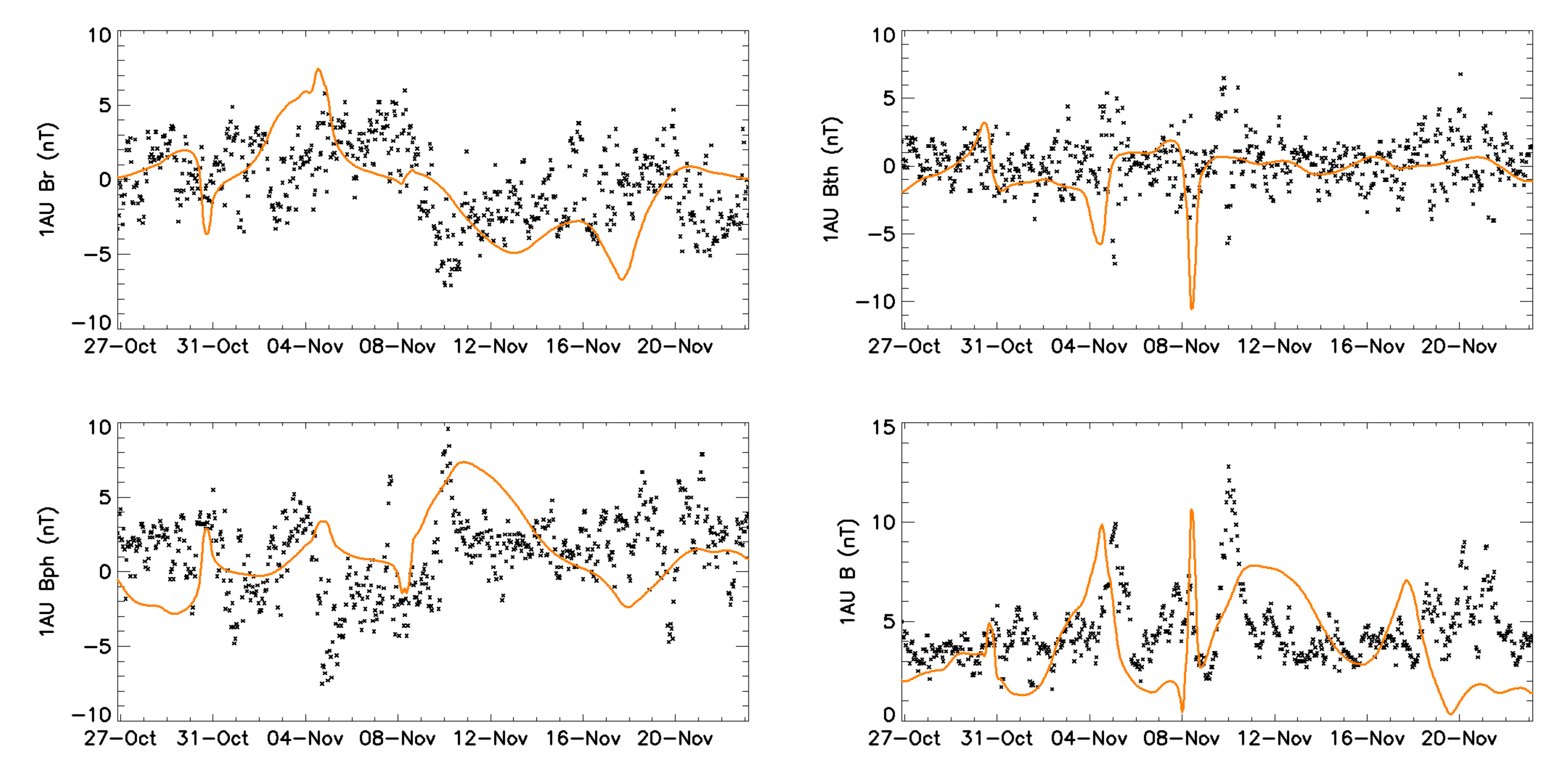}
\includegraphics[width=1\textwidth,clip=]{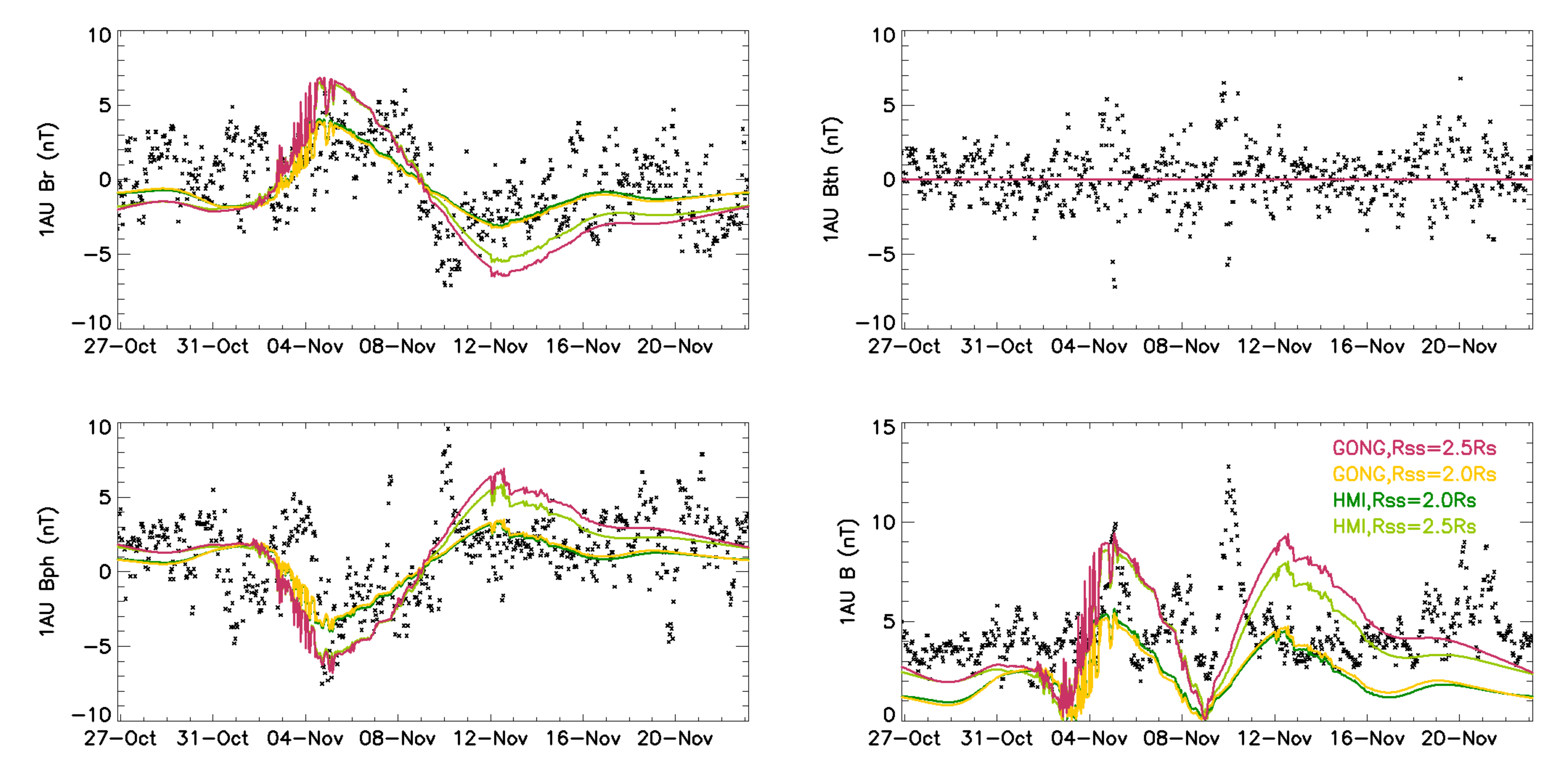}
  }
\caption{Comparison of MHD (orange curves in top two rows) and PFSS (colored curves in bottom two rows) results with near-Earth magnetic field observation (black).}
\label{c1au}
\end{center}
\end{figure}

\begin{table}[htb]
\centering
\caption{Evaluation based on 1AU observation to MHD and PFSS model. From left to right are successively the model, scaling coefficient, the polarity coincidence rate of three components with observed data, and the RMSE of scaled $B_r$, $B_\theta$, $B_\phi$ and $B$ relative to observed data.}
\label{tab:1AU}
\begin{tabular}{ccccccccc}
\toprule
Model&Scale&$P_r$&$P_\theta$&$P_\phi$&RMSE($B_r$)&RMSE($B_\theta$)&RMSE($B_\phi$)&RMSE($B$)\\
\midrule
MHD                           &4.40752&0.647866&0.548781&0.565549&3.23861&2.40222&3.56879&2.72186\\
PFSS (HMI, 2.5Rs)    &8.75295&0.682927& -             &0.772866&2.90104&1.77866&2.91423&2.54280\\
PFSS (HMI, 2.0Rs)    &3.25737&0.696646& -             &0.777439&2.33263&1.77866&2.52706&3.00398\\
PFSS (GONG, 2.0Rs)&3.68674&0.695122& -             &0.778963&2.34179&1.77866&2.51460&2.97959\\
PFSS (GONG, 2.5Rs)&10.8234&0.689024& -             &0.769817&3.18623&1.77866&3.11440&2.68109\\
\bottomrule
\end{tabular}
\end{table}

The results of LINFF's 3rd, 4th, 13th, and 14th simulations and AWSoM were compared with near-Earth magnetic field observation. The results are shown in Figure \ref{c1au} and Table \ref{tab:1AU}. The scaling factors determined earlier from PSP radial magnetic field measurement are still used here, and the solar wind velocity in PFSS model is obtained by slightly smoothing the PSP observation. It can be seen that PFSS model still has a relatively better performance in polarity prediction.

\subsection{The Role of PSP Near-Solar Observation Data}

\begin{figure}[htbp]
\begin{center}
\includegraphics[width=1\textwidth]{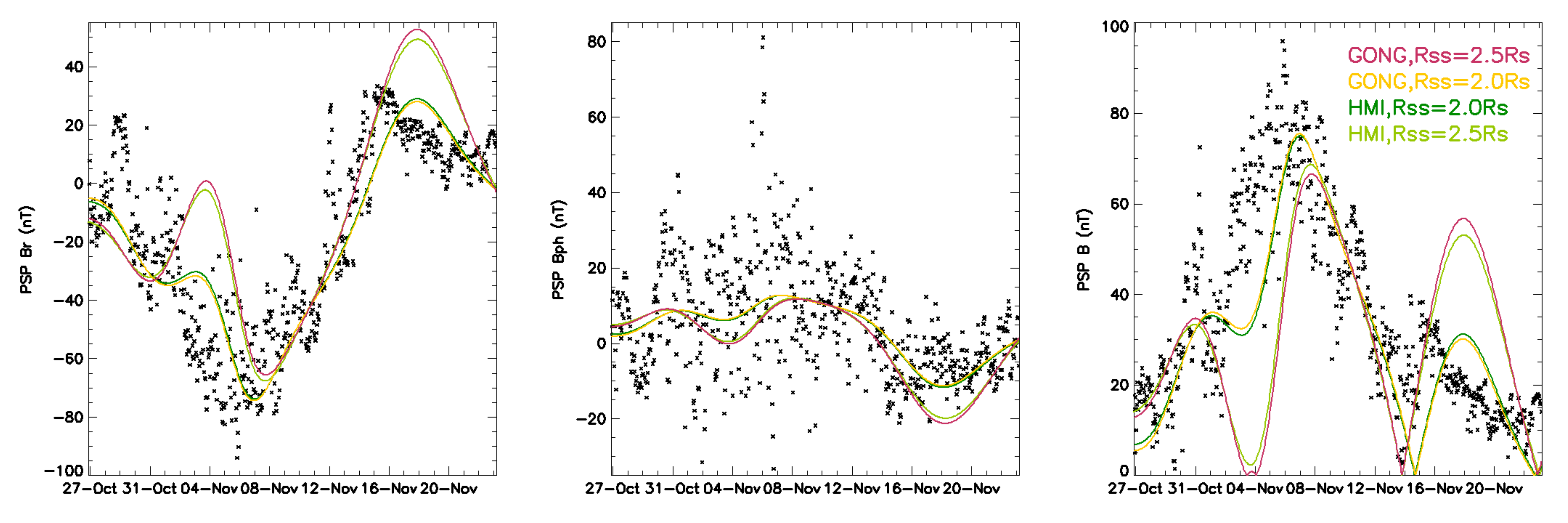}
\caption{Comparison of PFSS model results using constant solar wind velocity with PSP in-situ observation (black).}
\label{avg1}
\end{center}
\end{figure}

\begin{table}[htb]
\centering
\caption{Evaluation of PFSS model using constant solar wind velocity. From left to right are successively the model setting (magnetogram and the position of source surface), the scaling coefficient, the polarity coincidence rate of $B_r$ and $B_\phi$ components with observed data, and the RMSE of scaled $B_r$, $B_\phi$ and $B$ relative to observed data.}
\label{tab:avg1}
\begin{tabular}{ccccccc}
\toprule
PFSS&Scale&$P_r$&$P_\phi$&RMSE($B_r$)&RMSE($B_\phi$)&RMSE($B$)\\
\midrule
HMI, 2.5Rs    &10.1454&0.907012&0.769817&25.4906&14.2319&24.8593\\
HMI, 2.0Rs    &3.54765&0.882622&0.766768&17.0102&12.9171&14.2442\\
GONG, 2.0Rs&3.99926&0.881098&0.762195&16.8224&12.8831&13.9287\\
GONG, 2.5Rs&12.5992&0.878049&0.762195&26.7738&14.4556&26.2948\\
\bottomrule
\end{tabular}
\end{table}

In order to find out how much role the solar wind velocity measured by PSP plays in PFSS model, the potential field solutions obtained by the 3rd, 4th, 13th and 14th simulations in LINFF are extrapolated to interplanetary space with a constant radial solar wind velocity of 400 km/s and compared with the in-situ magnetic field measurement of PSP. The results are shown in Figure \ref{avg1} and Table \ref{tab:avg1}. As expected, the measured velocity improves the simulation of magnetic field strength and variation considerably.

\subsection{About the Near-Real-Time GONG Synoptic Magnetograms}

\begin{figure}[htbp]
\begin{center}
\vbox{
\includegraphics[width=1\textwidth,clip=]{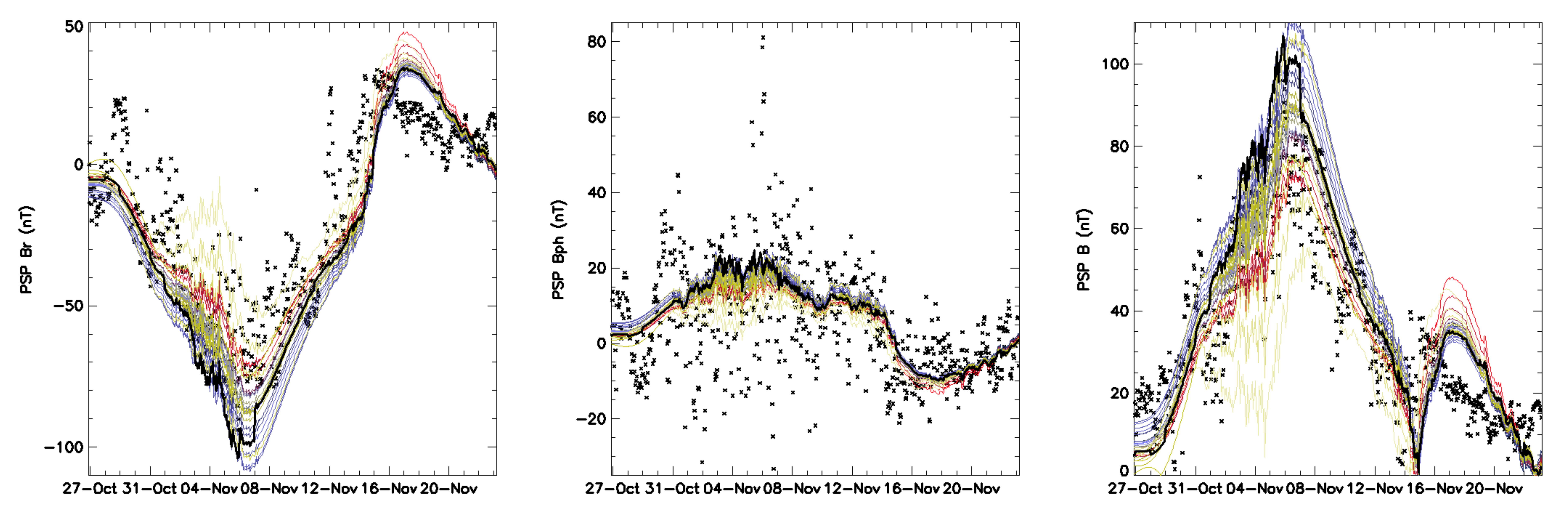}
\includegraphics[width=1\textwidth,clip=]{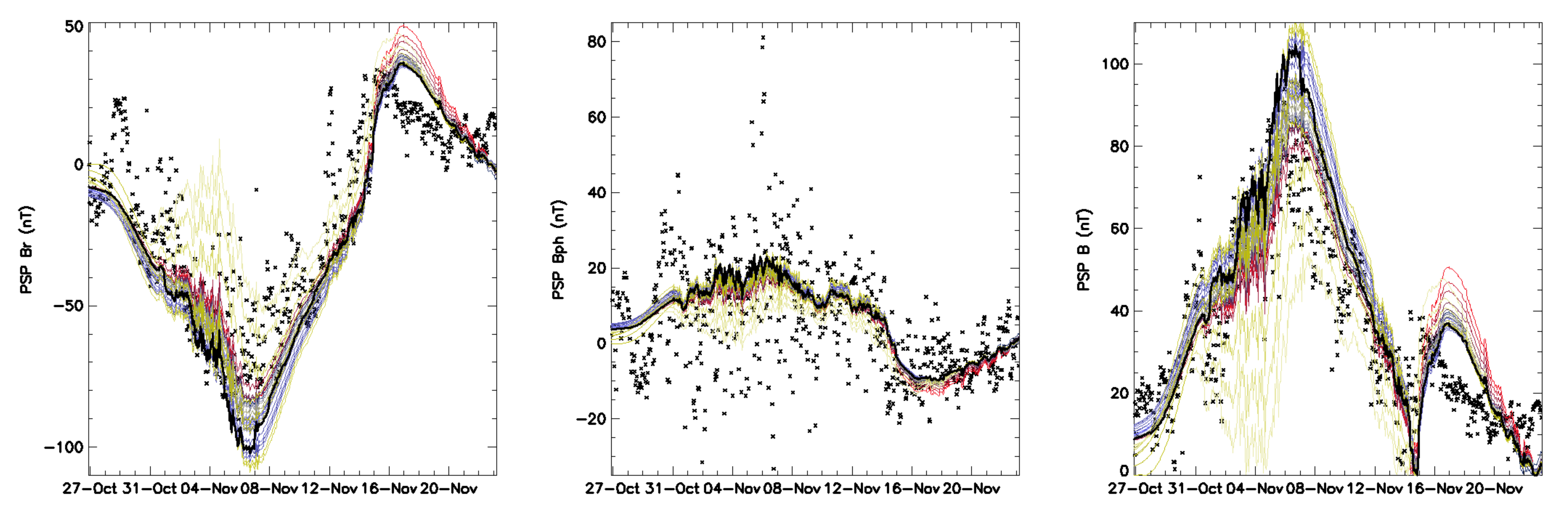}
\includegraphics[width=1\textwidth,clip=]{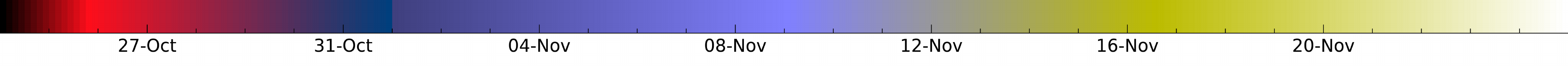}
  }
\caption{Magnetic field in CR2210 obtained by PFSS model with time-varying synoptic maps as input. The upper and lower rows are calculated with FDIPS and pfsspy respectively. Each colored curve is associated with an individual magnetogram while the black curve is spliced according to the principle of time proximity. The black dots are still from PSP observation.}
\label{multi2210}
\end{center}
\end{figure}

\begin{table}[htb]
\centering
\caption{Evaluation of multi-magnetogram PFSS model (CR2210). From left to right are successively the model, the scaling coefficient, the polarity coincidence rate of $B_r$ and $B_\phi$ components with observed data, and the RMSE of scaled $B_r$, $B_\phi$ and $B$ relative to observed data.}
\label{tab:multi2210}
\begin{tabular}{cccccccc}
\toprule
Model&Scale&$P_r$&$P_\phi$&RMSE($B_r$)&RMSE($B_\phi$)&RMSE($B$)\\
\midrule
PFSS (FDIPS)&6.95485&0.882622&0.757622&20.0187&13.1755&14.7814\\
PFSS (pfsspy)&5.35645&0.878049&0.762195&19.8198&12.9848&14.2208\\
\bottomrule
\end{tabular}
\end{table}

\begin{figure}[htbp]
\begin{center}
\includegraphics[width=1\textwidth]{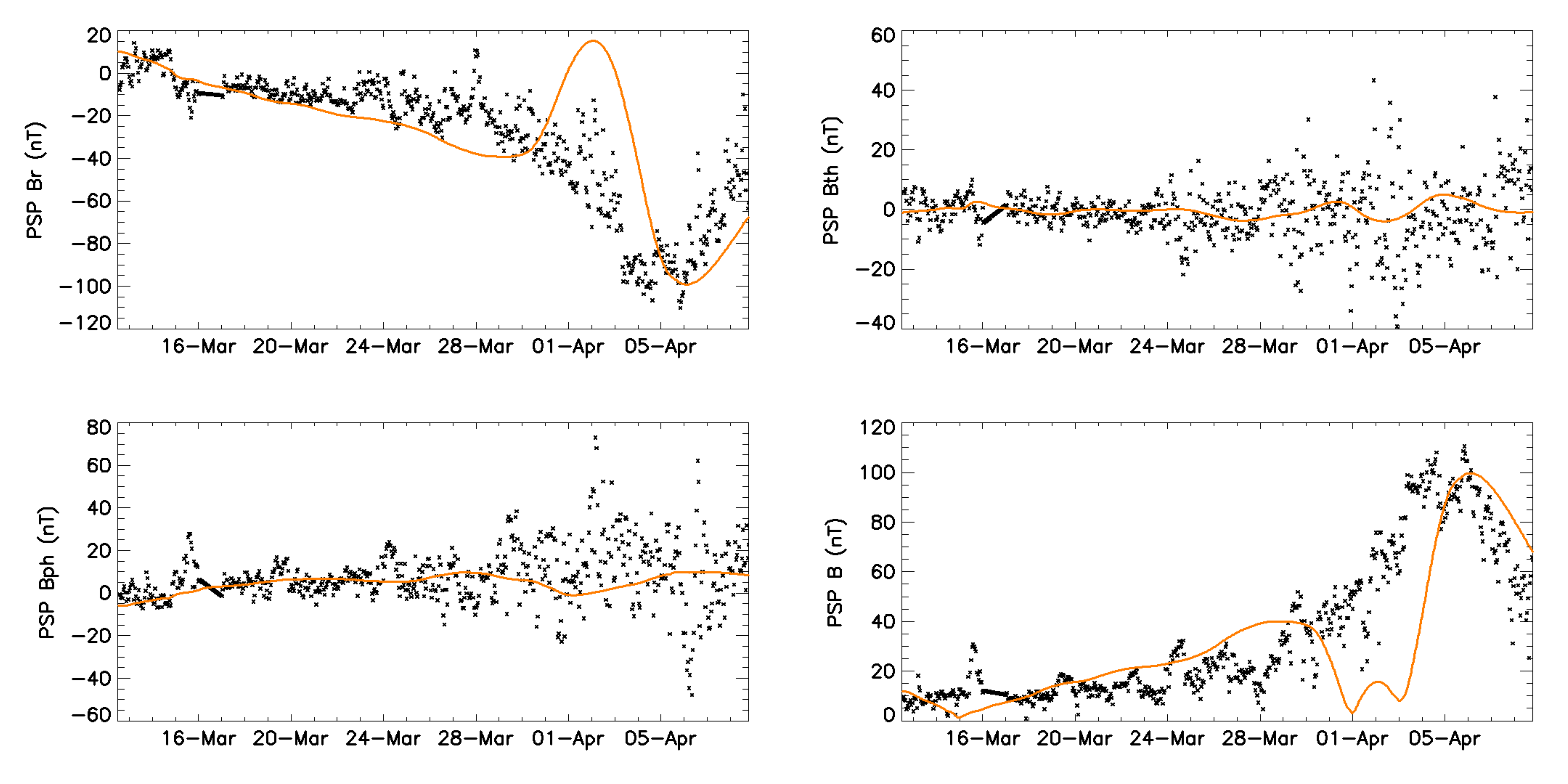}
\caption{The comparison of MHD model magnetic field (orange) and PSP in-situ data (black) in CR2215.}
\label{mhd2215}
\end{center}
\end{figure}

\begin{figure}[htbp]
\begin{center}
\vbox{
\includegraphics[width=1\textwidth,clip=]{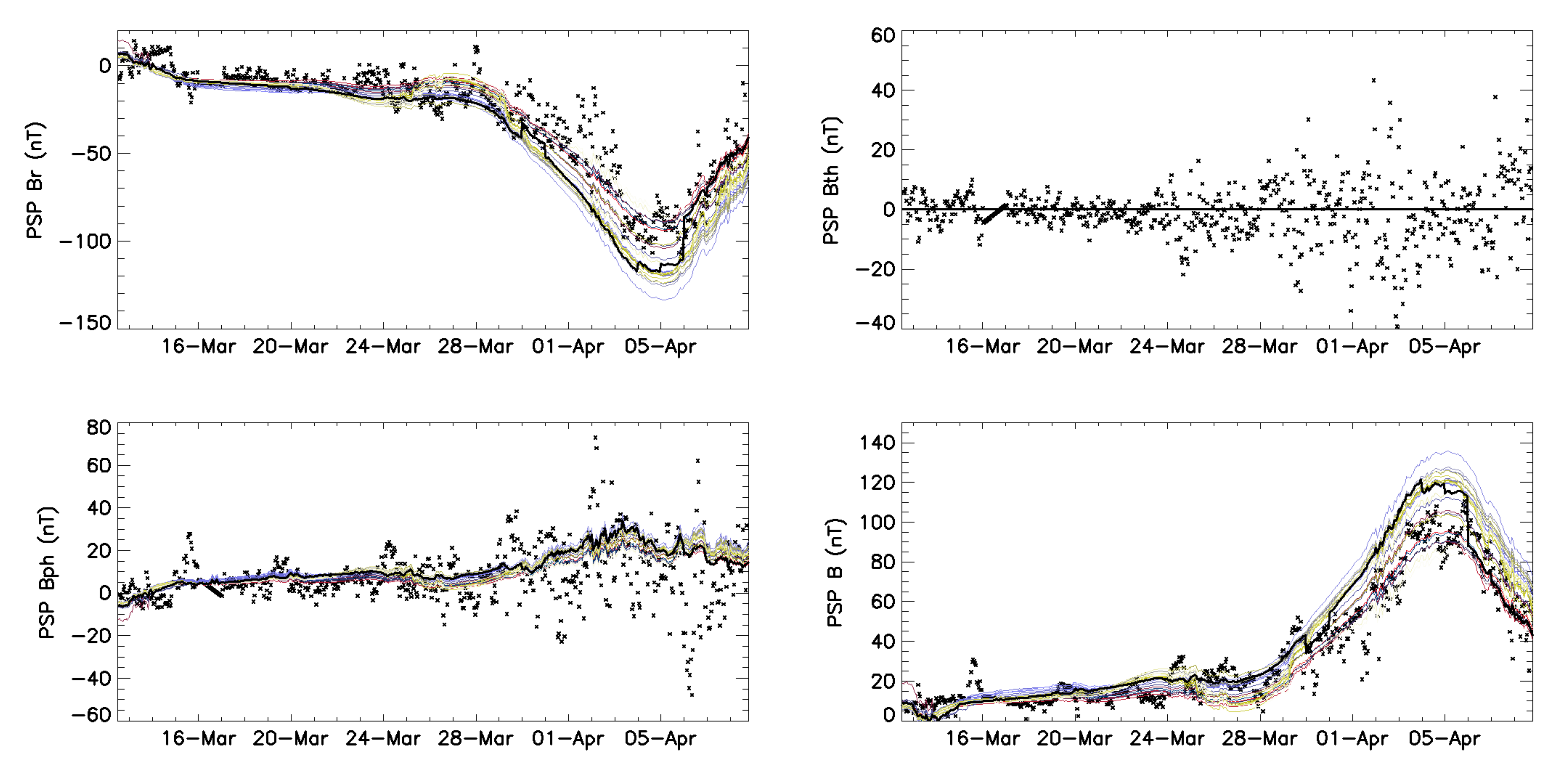}
\includegraphics[width=1\textwidth,clip=]{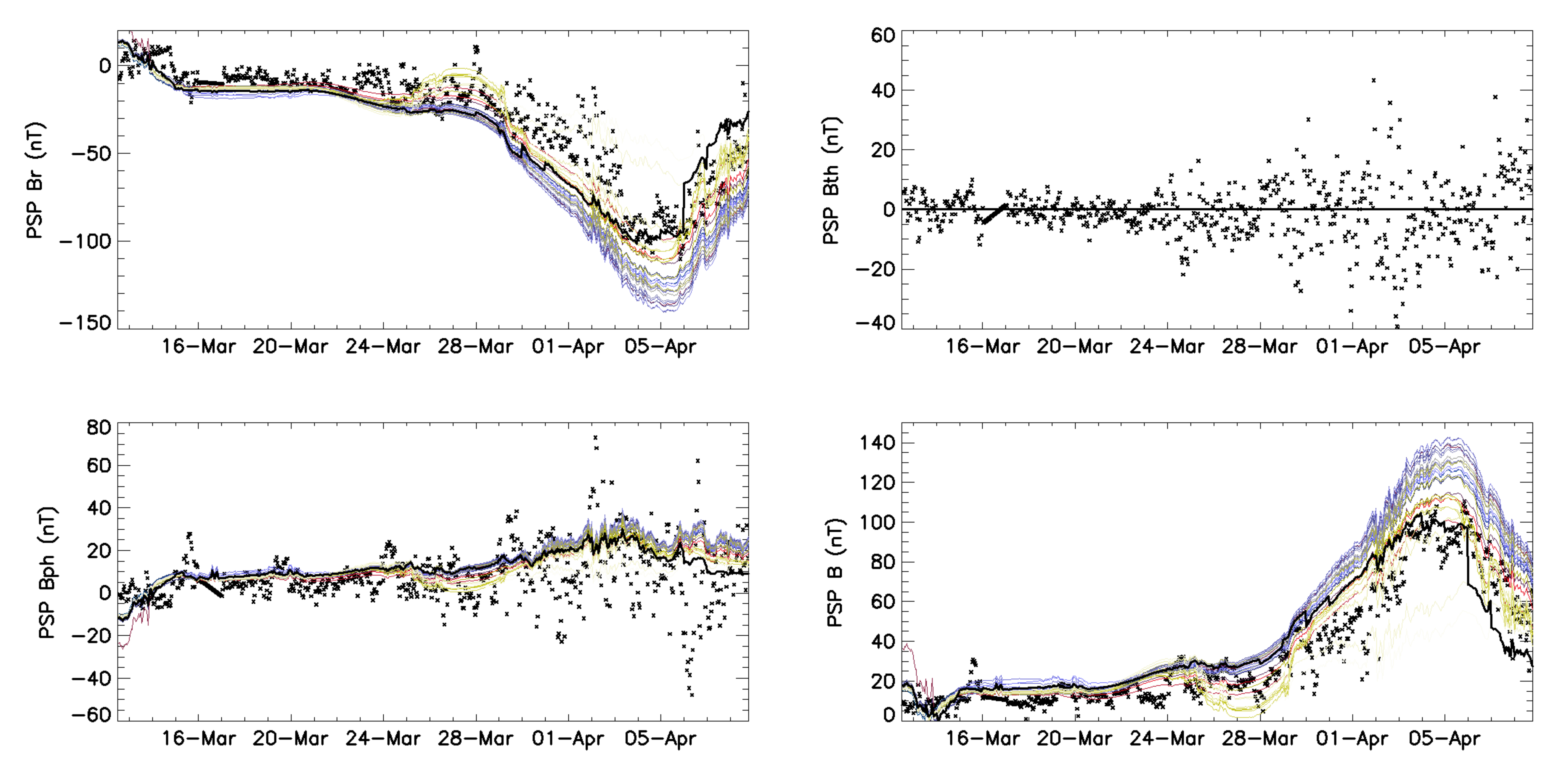}
\includegraphics[width=1\textwidth,clip=]{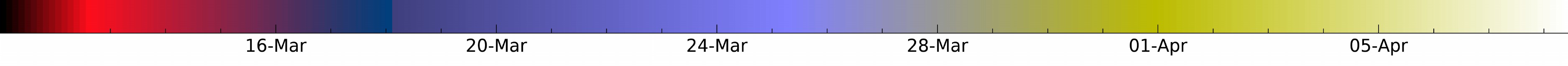}
  }
\caption{Magnetic field in CR2215 obtained by PFSS model with time-varying synoptic maps as input. The upper two and lower two rows are calculated with FDIPS and pfsspy respectively. Each colored curve is associated with an individual magnetogram while the black curve is spliced according to the principle of time proximity. The black dots are still from PSP observation.}
\label{multi2215}
\end{center}
\end{figure}

\begin{table}[htb]
\centering
\caption{Evaluation of MHD and multi-magnetogram PFSS model (CR2215). From left to right are successively the model, scaling coefficient, the polarity coincidence rate of three components with observed data, and the RMSE of scaled $B_r$, $B_\theta$, $B_\phi$ and $B$ relative to observed data.}
\label{tab:multi2215}
\begin{tabular}{ccccccccc}
\toprule
Model&Scale&$P_r$&$P_\theta$&$P_\phi$&RMSE($B_r$)&RMSE($B_\theta$)&RMSE($B_\phi$)&RMSE($B$)\\
\midrule
MHD&2.45207&0.885496&0.496183&0.770992&25.7666&9.63254&13.9690&23.0783\\
PFSS (FDIPS)&9.72423&0.938931&-&0.787786&16.1711&9.40109&13.6049&13.6724\\
PFSS (pfssspy)&13.2842&0.940458&-&0.789313&17.1665&9.40109&13.3503&14.7113\\
\bottomrule
\end{tabular}
\end{table}

Considering the high temporal resolution of GONG synoptic maps, it could be used to further optimize the PFSS model and there has been some work done for it \cite{Badman}, so we tried to update the magnetic map once a day in CR2210, select the source surface at 2.0 solar radii, and recalculate the results of FDIPS and pfsspy using a grid of $150\times180\times360$, with Figure \ref{multi2210} and Table \ref{tab:multi2210} displaying the results. The same approach is applied to CR2215 to obtain Figure \ref{multi2215}, while we also present the results of MHD model as reference in Figure \ref{mhd2215} and evaluate their performance in Table \ref{tab:multi2215}. AWSoM parameters here are recommended values Poynting ratio $=1.0\times 10^6$ J/(m$^2$sT) and Coronal Heating $=1.5\times 10^5$ mT$^{1/2}$.

The analysis shows that updating the input over time does not seem to improve the overall performance of PFSS model as expected. This may be because the zero point corrected magnetograms have been individually scaled to varying degrees during generation and it may not be appropriate to combine them directly in this way. However, we can speculate that if GONG standard maps are used in model, the refinement of time mesh will have a good effect. In addition, we assume that the radial velocity of each flux tube remains constant over a Carrington rotation to use PSP solar wind velocity data, so it is theoretically more appropriate to use a constant photosphere magnetogram here.

\section{Conclusions \& discussion}
In this paper, we used in-situ measurements of solar wind around the first and second perihelion of  PSP to obtain coronal and interplanetary magnetic fields. Combining the Potential Field Source Surface (PFSS) coronal magnetic field model and the Parker spiral interplanetary magnetic field model is a common method to describe the solar-terrestrial space magnetic structure. An initial method for solving the potential field is by spherical harmonic functions, which provides the most accurate results for the model but the calculation process consumes a lot of time. Among those algorithms, compared with using the standard analytical form truncation or directly performing iterative difference calculation (such as FDIPS) on the Laplace equation, separating variables according to the structure of the analytical solution after constructing the difference scheme then converting to an eigenvalue problem (such as pfsspy) or using numerical methods to perform fast spherical harmonic transformation \cite{Suda} in the analytical process can often combining advantages of the previous two and significantly improve computational efficiency. An important parameter for the solution of spiral interplanetary magnetic field is solar wind speed, and after the launch of PSP, the actual measured data other than a constant can be used, which is of great significance to the prediction of space weather. We can also obtain coronal and interplanetary magnetic fields through MHD methods (such as AWSoM).

By comparing the simulated results with interplanetary magnetic field observed around PSP perihelions, we found that the measured solar wind velocity significantly improved the fitting effect of PFSS model. The source surface setting at 2.0Rs and 2.5Rs gave similar magnetic polarity predictions, but the former simulated magnetic field strength and variation better than the latter. The optimal value of source surface may be changing in different Carrington Rotation and under the contrast with different observational data, which needs further profound study. The interplanetary magnetic field intensity obtained based on HMI magnetogram is higher than that based on GONG's and the result is further improved with mesh density increasing. However, the performance of GONG magnetogram is more stable under sparse grids. GONG synoptic maps has a temporal resolution of 1 hour or so, which makes it possible to conduct more reliable magnetic field predictions by continuously updating the input, but this method should be prudently applied to the zero point corrected products and it is better combined with the evolution of velocity field. Combining the PSP's velocity observation with those of other spacecrafts might be able to implement that, and if a larger range of velocity field could be built, it could be used to optimize the simulation of near-earth magnetic field.

In the basic PFSS model, the $B_\theta$ component is always zero outside the source surface. Although this may not seem unreasonable because the measured magnetic field is relatively weak, it can be further improved by certain methods. For example, the source surface can be set to a non-spherical or non-heliocentric shape. Potential field can also be combined with local magnetic field modeling such as heliospheric current sheets, solar active regions, and coronal mass ejections. In-situ observation of interplanetary space magnetic field can also provide constraints for Parker spiral. MHD simulations, while providing more information about the longitudinal magnetic field, are not reliable enough to predict the direction and magnitude, which can be similarly optimized. Moreover, although measurement of the transverse magnetic field of solar photosphere are not that accurate at present, it might be helpful to include it as input data to the models as well.

The ``open flux problem'' is a long-standing but unsolved problem in the coronal and interplanetary magnetic field modeling, which yield magnetic field lower than in-situ observations. Scaling results with a fixed coefficient works well for PFSS models, but may not suitable for MHD models. The problem of underestimation may stem from the inaccuracy of existing polar magnetic field measurements. We look forward to more precise polar measurements by the Solar Orbiter mission, enabling the construction of more frequent and accurate synoptic maps for model input. On the other hand, we will continue to optimize various magnetic field models, hoping that through the improvement of both the model and observation, the ``open flux problem'' can finally be solved.


\begin{thebibliography}{20}
\providecommand\natexlab[1]{#1}
\providecommand\JournalTitle[1]{#1}
\bibitem[{Badman} {et~al.}(2020)]{Badman}
{Badman}, S.~T., {Bale}, S.~D., {Mart{\'\i}nez Oliveros}, J.~C., {et~al.} 2020,
  Astrophysical Journal Supplement Series, 246, 23
\bibitem[{Bale} {et~al.}(2016)]{Bale2016}
{Bale}, S.~D., {Goetz}, K., {Harvey}, P.~R., {et~al.} 2016, Space Science
  Reviews, 204, 49
\bibitem[{Fox} {et~al.}(2016)]{Fox2016}
{Fox}, N.~J., {Velli}, M.~C., {Bale}, S.~D., {et~al.} 2016, Space Science
  Reviews, 204, 7
\bibitem[{Kasper} {et~al.}(2016)]{Kasper2016}
{Kasper}, J.~C., {Abiad}, R., {Austin}, G., {et~al.} 2016, Space Science
  Reviews, 204, 131
\bibitem[Li {et~al.}(2021)]{Li2021}
Li, H., Feng, X., \& Wei, F. 2021, Journal of Geophysical Research: Space
  Physics, 126, e2020JA028870, e2020JA028870 2020JA028870
\bibitem[{Linker} {et~al.}(2017)]{Linker}
{Linker}, J.~A., {Caplan}, R.~M., {Downs}, C., {et~al.} 2017, Astrophysical
  Journal, 848, 70
\bibitem[{Miki{\'c}} {et~al.}(2018)]{Mikic2018}
{Miki{\'c}}, {}, Z., {Downs}, C., {et~al.} 2018, Nature Astronomy, 2, 913
\bibitem[{Parker}(1958)]{Parker1958}
{Parker}, E.~N. 1958, Astrophysical Journal, 128, 664
\bibitem[{Riley} {et~al.}(2019)]{Riley}
{Riley}, P., {Linker}, J.~A., {Mikic}, Z., {et~al.} 2019, Astrophysical
  Journal, 884, 18
\bibitem[{Ruan} {et~al.}(2008)]{Ruan2008}
{Ruan}, P., {Wiegelmann}, T., {Inhester}, B., {et~al.} 2008, Astronomy and
  Astrophysics, 481, 827
\bibitem[{Schatten} {et~al.}(1969)]{Pfss}
{Schatten}, K.~H., {Wilcox}, J.~M., \& {Ness}, N.~F. 1969, Solar Physics, 6,
  442
\bibitem[{Suda} \& {Takami}(2002)]{Suda}
{Suda}, R., \& {Takami}, M. 2002, Mathematics of Computation, 71, 703
\bibitem[T{\'{o}}th {et~al.}(2011)]{Toth}
T{\'{o}}th, G., van~der Holst, B., \& Huang, Z. 2011, Astrophysical Journal,
  732, 102
\bibitem[van~der Holst {et~al.}(2014)]{Van}
van~der Holst, B., Sokolov, I.~V., Meng, X., {et~al.} 2014, Astrophysical
  Journal, 782, 81
\bibitem[Wiegelmann(2011)]{Linff}
Wiegelmann, T. 2011, LINFF: An IDL-widget program for force-free coronal
  magnetic fields., 5th edn., Max-Planck-Institut f$\ddot{u}$r Sonnensystemforschung,
  Max-Planck-Strasse 2, 37191 Katlenburg-Lindau, Germany
\bibitem[{Wiegelmann} {et~al.}(2020)]{Wiegelmann2020}
{Wiegelmann}, T., {Neukirch}, T., {Nickeler}, D.~H., \& {Chifu}, I. 2020, Solar
  Physics, 295, 145
\bibitem[{Wiegelmann} \& {Sakurai}(2021)]{Wiegelmann2021}
{Wiegelmann}, T., \& {Sakurai}, T. 2021, Living Reviews in Solar Physics, 18, 1
\bibitem[{Yang} {et~al.}(2020{\natexlab{a}})]{YangZH2020b}
{Yang}, Z., {Tian}, H., {Tomczyk}, S., {et~al.} 2020{\natexlab{a}}, Science in
  China E: Technological Sciences, 63, 2357
\bibitem[{Yang} {et~al.}(2020{\natexlab{b}})]{YangZH2020a}
{Yang}, Z., {Bethge}, C., {Tian}, H., {et~al.} 2020{\natexlab{b}}, Science,
  369, 694
\bibitem[Yeates(2020)]{Pfsspy}
Yeates, A.~R. 2020, Notes on the Numerical Methods in pfsspy, Department of
  Mathematical Sciences, Durham University, UK
\end{thebibliography}
\end{document}